\documentclass[twocolumn,prx,9pt,superscriptaddress,amsfonts]{revtex4-1}

\usepackage{graphicx}
\usepackage{hyperref} 
\usepackage{color}
\usepackage{bm}
\usepackage{mathtools}
\usepackage{graphicx}
\usepackage{graphics}
\usepackage{dcolumn}
\usepackage{float}
\usepackage{epsfig}
\usepackage{epstopdf} 
\usepackage{color}
\usepackage{amsmath}
\usepackage{textcomp}

\makeatletter
\renewcommand*{\@caption@fignum@sep}{ $|$}
\makeatother

\pdfpageattr {/Group << /S /Transparency /I true /CS /DeviceRGB>>} 

\begin{document}

\title{Controlling spin-orbit interactions in silicon quantum dots using magnetic field direction}

\author{Tuomo Tanttu}
\affiliation{Center for Quantum Computation and Communication Technology, School of Electrical Engineering and Telecommunications, The University of New South Wales, Sydney, NSW 2052, Australia.}
\author{Bas Hensen}
\affiliation{Center for Quantum Computation and Communication Technology, School of Electrical Engineering and Telecommunications, The University of New South Wales, Sydney, NSW 2052, Australia.}
\author{Kok Wai Chan}
\altaffiliation{current address: Microsoft Quantum, The University of Sydney, Sydney, NSW 2006, Australia.}
\affiliation{Center for Quantum Computation and Communication Technology, School of Electrical Engineering and Telecommunications, The University of New South Wales, Sydney, NSW 2052, Australia.}
\author{Chih Hwan Yang}
\affiliation{Center for Quantum Computation and Communication Technology, School of Electrical Engineering and Telecommunications, The University of New South Wales, Sydney, NSW 2052, Australia.}
\author{Wister Huang}
\affiliation{Center for Quantum Computation and Communication Technology, School of Electrical Engineering and Telecommunications, The University of New South Wales, Sydney, NSW 2052, Australia.}
\author{Michael Fogarty}
\altaffiliation{current address: London Centre for Nanotechnology, UCL, 17-19 Gordon St, London WC1H 0AH, United Kingdom}
\affiliation{Center for Quantum Computation and Communication Technology, School of Electrical Engineering and Telecommunications, The University of New South Wales, Sydney, NSW 2052, Australia.}
\author{Fay Hudson}
\affiliation{Center for Quantum Computation and Communication Technology, School of Electrical Engineering and Telecommunications, The University of New South Wales, Sydney, NSW 2052, Australia.}
\author{Kohei Itoh}
\affiliation{School of Fundamental Science and Technology, Keio University, 3-14-1 Hiyoshi, Kohokuku, Yokohama 223-8522, Japan.}
\author{Dimitrie Culcer}
\affiliation{School of Physics, The University of New South Wales, Sydney 2052, Australia.}
\affiliation{ARC Centre for Excellence in Future Low-Energy Electronics Technologies, Sydney 2052, Australia.}
\author{Arne Laucht}
\affiliation{Center for Quantum Computation and Communication Technology, School of Electrical Engineering and Telecommunications, The University of New South Wales, Sydney, NSW 2052, Australia.}
\author{Andrea Morello}
\affiliation{Center for Quantum Computation and Communication Technology, School of Electrical Engineering and Telecommunications, The University of New South Wales, Sydney, NSW 2052, Australia.}
\author{Andrew Dzurak}
\affiliation{Center for Quantum Computation and Communication Technology, School of Electrical Engineering and Telecommunications, The University of New South Wales, Sydney, NSW 2052, Australia.}

\begin{abstract}
Silicon quantum dots are considered an excellent platform for spin qubits, partly due to their weak spin-orbit interaction. However, the sharp interfaces in the heterostructures induce a small but significant spin-orbit interaction which degrade the performance of the qubits or, when understood and controlled, could be used as a powerful resource. To understand how to control this interaction we build a detailed profile of the spin-orbit interaction of a silicon metal-oxide-semiconductor double quantum dot system. We probe the derivative of the Stark shift, $g$-factor and $g$-factor difference for two single-electron quantum dot qubits as a function of external magnetic field and find that they are dominated by spin-orbit interactions originating from the vector potential, consistent with recent theoretical predictions. Conversely, by populating the double dot with two electrons we probe the mixing of singlet and spin-polarized triplet states during electron tunneling, which we conclude is dominated by momentum-term spin-orbit interactions that varies from 1.85~MHz up to 27.5~MHz depending on the magnetic field orientation. Finally, we exploit the tunability of the derivative of the Stark shift of one of the dots to reduce its sensitivity to electric noise and observe an 80~\% increase in~$T_2^*$. We conclude that the tuning of the spin-orbit interaction will be crucial for scalable quantum computing in silicon and that the optimal setting will depend on the exact mode of qubit operations used.
\end{abstract}

\maketitle

Silicon-based spin qubits have attracted attention as candidates for large scale quantum computing thanks to their long coherence times, excellent controllability and fabrication techniques that are well established in the semiconductor industry~\cite{Zwanenburg2013, Pla2012, Fuechsle2012, Veldhorst2015, Bienfait2015, Mi2016, Maurand2016, Fujita2017, Samkharadze2018, Watson2018, Zajac2018, Jock2018, Nakajima2018}. Even though being weak compared, for instance, to GaAs, the spin-orbit interaction (SOI) significantly affects the behaviour of silicon spin qubits, especially through the dependency of the SOI on the valley state~\cite{Jock2018, Veldhorst2015b, Ruskov2018}. SOI is responsible for effects such as the Stark shift of the electron spin resonance (ESR) frequency, variation of Lande $g$-factors, and mixing between singlet (S) and polarized triplet (T$^-$) states~\cite{Ferdous2018, Ferdous2018_2, Ruskov2018,Stepanenko2012}. These effects can be harnessed, for instance, to drive the ESR transition electrically via Stark shift or by exploiting the variation in the $g$-factors to address qubits individually with a global microwave (MW) field~\cite{Corna2018, Huang2017, Veldhorst2015b, Huang2018arxiv}. In contrast, spin-orbit effects such as spin-flip tunneling and strong Stark shift can cause state leakage or increased sensitivity to electric noise~\cite{Ferdous2018, Veldhorst2015b, Hofmann2017}. Hence, understanding and controlling the SOI will be important for spin qubit control in larger arrays of dots in the future~\cite{Jones2018, Fogarty2018, Veldhorst2017}.

\begin{figure*}
\centering
\includegraphics[width=\linewidth]{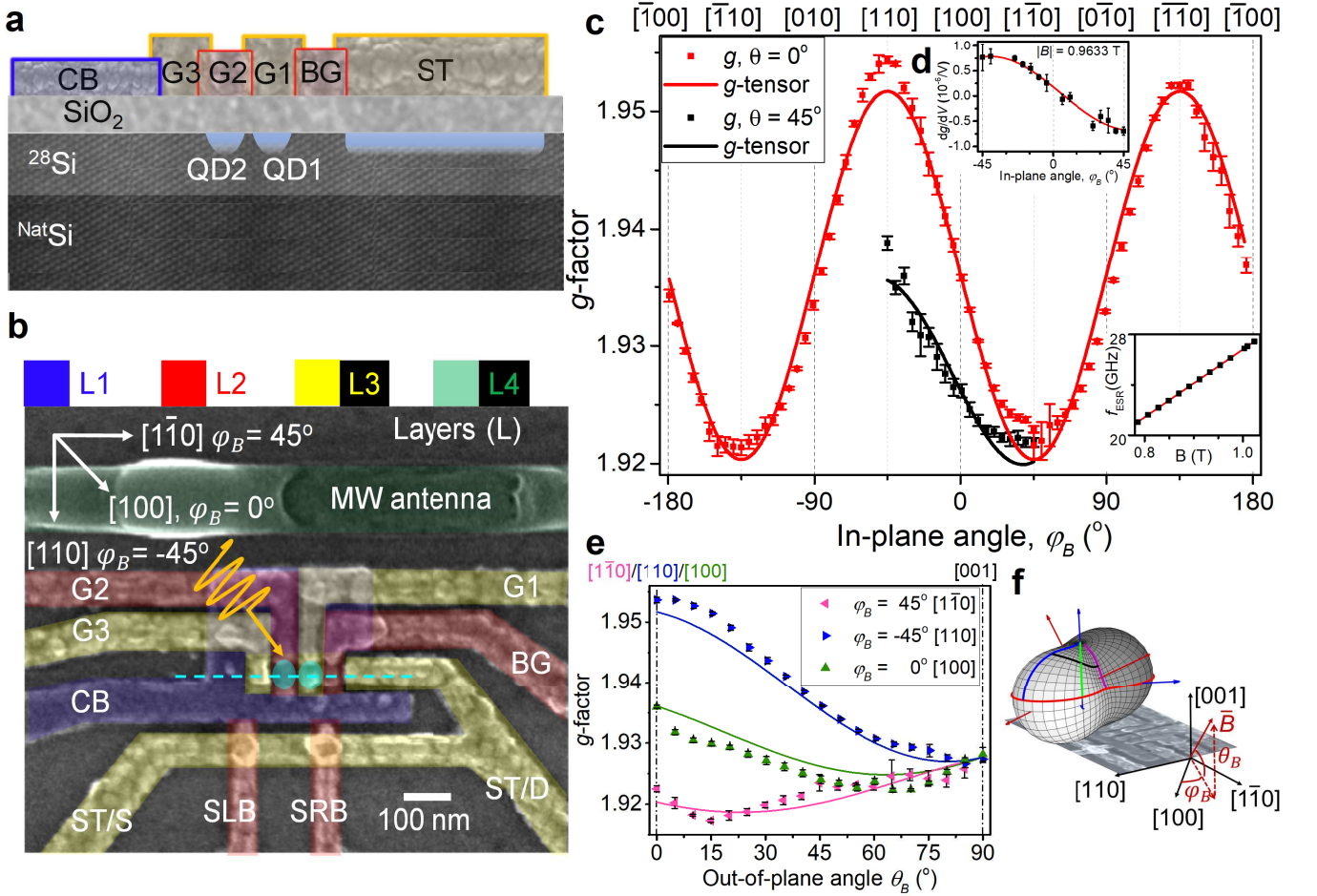}
\caption{Sample, $g$-factor measurement (a) Schematic cross section at the position of the dots (QD1 and QD2) along the dotted line in (b). (b) False colour SEM image of our sample. MW antenna used to drive the qubits is on the top, qubit dots in the middle, and SET used to sense the dots is at the bottom. (c) $g$-factor of the electron occupying the G1 dot as a function of external magnetic field angle in-plane and at 45$^\textrm{o}$ out-of-plane of the sample in the (1,0) charge configuration. The solid lines corresponds to the estimate extracted from the complete g-tensor. Inset: Example of ESR frequency as a function of magnetic field with a linear fit that is used to extract the g-factor. (d) The derivative of the Stark shift as a function of in-plane magnetic field angle. (e) Out-of-plane $g$-factors measured (coloured symbols) with estimates from complete $g$-tensor (solid lines). (f) Notation of the magnetic field angles with respect to the sample and isosurface of $g$-factor of the G1 dot based on a single $g$-tensor. We have subtracted 1.9 in order to visualize the anisotropy of the $g$-factor. Blue arrows correspond to laboratory coordinates and red arrows correspond to the principal axes of the $g$-factor ellipsoid. The data in (c) and (e) are taken along the lines shown on the surface.}\label{SEMs}
\end{figure*}

Here, we fully characterize the SOI and demonstrate how we can tune it in a silicon metal-oxide-semiconductor (SiMOS) double quantum dot (QD) structure. The structure studied here is shown in Fig.~\ref{SEMs}b and described in Ref.~\cite{Fogarty2018}. We vary the direction of the external magnetic field and measure quantities such as the $g$-factor of one dot, the $g$-factor difference between the dots, the Stark effect, the dephasing time $T_2^*$, and S-T$^-$ mixing. The $g$-factor, Stark shift and S-T$^-$ mixing exhibit sinusoidal depedence on the magnetic field direction, as reported before~\cite{Brauns2016, Jock2018, Ferdous2018_2, Ruskov2018}. We use these measurements to extract the Rashba and Dresselhaus interaction strengths of the lower energy valley state in both of the dots. By adjusting the magnetic field direction, the $g$-factor difference can be minimized allowing global ESR, or it can be maximized in order to address the qubits individually. We employ the tunability of the SOI to reduce the sensitivity of the dot to charge noise and observe an increase of 80\% in $T_2^*$ near the point where the derivative of the Stark shift vanishes~\cite{Ferdous2018, Ruskov2018}. Finally, from studying the S-T$^-$ transition we identify that the coupling in this system is caused by the dynamic spin-orbit field induced by charges moving between the dots, rather than by the hyperfine coupling or the differences in the $g$-tensors between the dots~\cite{Stepanenko2012, Stepanenko2003}. This answers a question raised in a previous study~\cite{Fogarty2018}. By adjusting the magnetic field direction we can minimize the mixing between the spin states. This could be extremely useful in reducing errors during spin shuttling of electron spins in a quantum bus, or in reducing undesired leakage to the T$^-$ state in the S-T operational basis.


\section{Spin-orbit interaction in silicon quantum dots}

Three main mechanisms are responsible for SOI in QDs: structural inversion asymmetry (Rashba)~\cite{Rashba1960}, bulk inversion asymmetry (Dresselhaus)~\cite{Dresselhaus1955}, and interface inversion asymmetry~\cite{Golub2004}. In bulk silicon the Dresselhaus term is absent, however as shown in Ref.~\citenum{Golub2004}, interface inversion asymmetry has the same representation in the Hamiltonian as the Dresselhaus term. Hence to keep the terminology simple, we refer to this as the Dresselhaus term. In a quantum dot the Rashba interaction leads to a renormalization of the in-plane $g$-factor, while the Dresselhaus interaction gives rise to shear terms in the in-plane Zeeman response leading to an anisotropy in the $g$-factor~\cite{Veldhorst2015b, Ruskov2018, Ferdous2018, Ferdous2018_2, Jock2018}. These effects allow tuning of the SOI by changing the orientation of the external magnetic field. Since at silicon interfaces the Dresselhaus term is expected to be dominant~\cite{Ferdous2018, Ferdous2018_2, Ruskov2018}, we can completely turn off SOI on demand by inducing a Dresselhaus effect that is equally strong but opposite in sign to the Rashba effect~\cite{Ferdous2018_2, Nestoklon2008, Ruskov2018, Jock2018, Hofmann2017}.

In silicon, a spin-orbit (SO) Hamiltonian for the $i$th valley can be written as
\begin{equation}
H_{\textrm{SO},v_i} = \alpha_i(k_x\sigma_y-k_y\sigma_x) + \beta_i(k_x\sigma_x-k_y\sigma_y), i= 1,2,\label{SOHamiltonian}
\end{equation}
where $\alpha$ and $\beta$ are the Rashba and Dresselhaus interaction coefficients, $k_x$ and $k_y$ are the electron wave operators along [100] and [010] lattice directions respectively, and $\sigma_x$ are $\sigma_y$ are the Pauli spin matrices. The electron wave operator is represented as $k_x = -i\frac{d}{dx}+eA_x/\hbar$, where $A_x$ is the $x$-component of the vector potential of the magnetic field, $e$ is the elementary charge and $\hbar$ is the reduced Planck's constant. In Ref.~\cite{Ruskov2018} it is shown that by choosing a gauge $\overrightarrow{A} = (B_y z,-B_x z,0)$ and by averaging over the $z$-axis we obtain from Eq.~\ref{SOHamiltonian}. corrections to the $g$-tensor. Here, we use two different experimental spin-orbit coefficients to describe the system. We use $\alpha_g$ and $\beta_g$ for the Rashba and Dresselhaus coefficients that we associate with the vector potential contribution (namely $g$-factor and Stark shift). Secondly, two dot quantities that are affected by the $\frac{d}{dx}$ term in $H_{\textrm{SO}}$ (namely S-T$^-$ mixing) we associate coefficients $\alpha_t$ and $\beta_t$. 

The $g$-tensor of a silicon quantum dot at an interface can be expressed as a 3$\times$3 matrix and it usually assumes the expression
\begin{equation}
\hat{g} = \begin{pmatrix}
     g_\parallel-\frac{2\alpha_g^*}{\mu_B} & \frac{2\beta_g^*}{\mu_B} & g_{xz}   \\
    \frac{2\beta_g^*}{\mu_B} & g_\parallel-\frac{2\alpha_g^*}{\mu_B} & g_{yz}   \\
    g_{xz} & g_{yz} & g_\perp
   \end{pmatrix},\label{Gtensor}
\end{equation}
where $g_\parallel$ is the $g$-factor in-plane with the quantum dot, $g_\perp$ is the $g$-factor perpendicular to the quantum dot plane, $\mu_B$ is the Bohr magneton and $\alpha_g^* = \frac{e\langle z-z_i\rangle\alpha_g}{\hbar}$, $\beta_g^*=\frac{e\langle z\rangle\beta_g}{\hbar}$, with $\langle z-z_i\rangle$ being the spread of the electron wave function in the $z$-direction ($z_i$ is the location of the interface)~\cite{Veldhorst2015b, Ruskov2018, Jock2018, Crippa2018}. Non-zero $g_{xz}$ and $g_{yz}$ terms can be generated from dipole matrix elements~\cite{Yang2013} for example from strong in-plane electric fields due to strain caused by thermal expansion mismatch when the device is cooled down~\cite{Pla2018}. For our quantum dots we use $\langle z\rangle = 1.68$~nm~\cite{Veldhorst2015b}. In silicon QDs, this tensor is diagonal and both $g_\parallel$ and $g_\perp$ are close to the vacuum value of the electron $g$-factor due to the large band gap in silicon~\cite{Roth1960}. Here, we assume a symmetric $g$-tensor and that the in-plane $B_x$ coordinate is aligned with the [100] crystal lattice direction. To obtain the $g$-factor along a certain direction, one can use $g = \sqrt{\hat{r}(\varphi_B,\theta_B)^\dagger \hat{g}_\textrm{1}^\dagger \hat{g}_\textrm{1} \hat{r}(\varphi_B,\theta_B)}$, where $\hat{r}(\varphi_B,\theta_B)$ is a unit vector pointing to the same direction as the external magnetic field $B$ expressed in Cartesian coordinates.

\section{Experimental $g$-tensor}

Fig.~\ref{SEMs}b shows the scanning electron microscope (SEM) image of the SiMOS device used in this experiment. This sample was previously used for experiments described in Ref.~\cite{Fogarty2018} and the structure is considered a candidate for scaling up to a logical qubit in silicon~\cite{Jones2018}. One quantum dot is induced at the Si/SiO$_2$ interface under each of the plunger gates G1 and G2, as shown in the schematic cross section in Fig.~\ref{SEMs}a. The quantum dots are confined laterally by a confinement barrier (CB) and a barrier gate (BG) which is also used to tune the tunnel rates between the reservoir and the dots. The reservoir is induced under an extension of our sensor top gate (ST). The MOS single-electron transistor (SET) sensor itself has two gates left and right (SLB, SRB) that are used to form barriers between the SET leads and the island which is capacitively coupled to the qubit dots in order to sense changes in the charge state. The MW antenna is used to coherently control the state of the qubit. We note from the image there is a discontinuity in the MW antenna, likely caused by an electric shock to the device. Despite this, we are able to drive the spin transition by applying MW frequency excitation to the antenna. We believe that this spin drive is either caused by electrical drive via valley mixing~\cite{Huang2017, Corna2018} or residual magnetic field drive despite the break or both~\cite{Muhonen2014}. We note that the extracted spin resonance frequency ($f_{\textrm{ESR}}$) does not depend on the driving mechanism.

In Fig.~\ref{SEMs}f we define the magnetic field direction with spherical coordinates so that $\varphi_B=0^\textrm{o}$ corresponds to the [100] Miller-index direction. This is tilted by 45 degrees from the main axis of the sample and the coils of the vector magnet that are aligned along the [110] lattice direction (corresponding to $\varphi_B = -45^\textrm{o}$). We also define $\theta_B=0^\textrm{o}$ when the magnetic field is in-plane with the sample and $\theta_B = 90^\textrm{o}$ when the magnetic field is pointing perpendicular to the sample plane and aligned with~[001].

The g-factor of QD1 for a particular external magnetic field direction $( \varphi_B, \theta_B )$ is determined in the (1,0) charge configuration by measuring $f_\textrm{ESR}$ as a function of magnetic field amplitude $|B|$. We obtain $g(\varphi_B,\theta_B)$ by fitting the linear slope (see inset of Fig.~\ref{SEMs}c) in order to exclude magnetic field hysteresis of the superconducting magnet coils. We show the measurements taken with the magnetic field in-plane with the sample in Fig.~\ref{SEMs}c together with measurements taken at 45 degrees out-of-plane. The out-of-plane datasets are shown in Fig.~\ref{SEMs}e. Initialization is performed by loading a spin-down electron while readout the spin state is based on spin dependent tunneling at the (0,0)$\rightarrow$(1,0) charge transition, using standard Elzerman readout~\cite{Elzerman2004}. To speed up the measurements, we employ an adiabatic ESR drive~\cite{Feher1956} to reduce the number of measurement points required to find $f_\textrm{ESR}$ down to 100~kHz accuracy using a constant 500~$\mu$s ESR pulse. 

\begin{figure*}
\centering
\includegraphics[width=\textwidth]{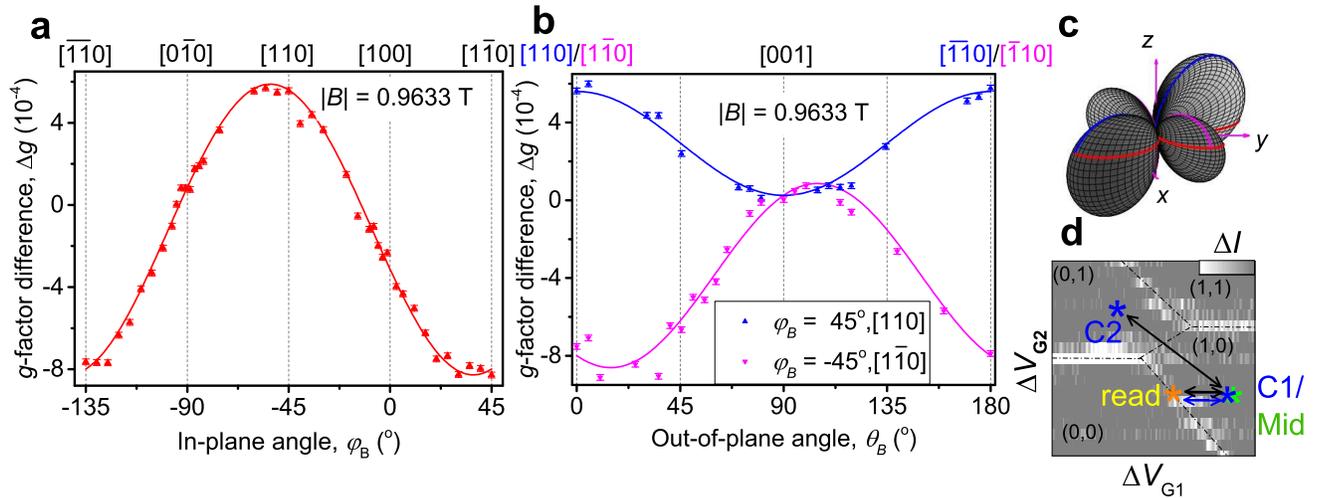}
\caption{$g$-factor difference between the dots. (a) Difference of $g$-factors between the dots under G1 and G2 with occupation (1,0) and (0,1) as a function of in-plane magnetic field angle together with sinusoidal fit. (b) Difference of $g$-factors between the dots under G1 and G2 as a function of out-of-plane magnetic field angle along [110] (blue) and [1$\bar{1}$0] (magenta). (c) Isosurface representation of the absolute value of $g$-factor difference as a function of magnetic field orientation. The magenta arrows represents crystal lattice orientation. Red, blue and magenta curves correspond to the fits in (a) and (b). (d) Pulse sequences (black and blue arrows) used to measure the $g$-factor difference in (a) and (b). C1 and C2 indicate the two different control points in the two different sequences.}\label{difference}
\end{figure*}

We use both the out-of-plane and in-plane data to fit a single symmetric $\hat{g}_1$ tensor for QD1. This $g$-tensor is
\begin{equation}
\hat{g}_{1} = 1.9\times I+
    \begin{pmatrix}
     36.0& -15.7& -5.7& \\
    -15.7& 36.0& -0.3&  \\
    -5.7& -0.3& 28.0&
    \end{pmatrix}\times 10^{-3},\label{GtensorQD1}
\end{equation}
where $I$ is the $3\times 3$ identity matrix. All the terms above have error bars of $\pm 10^{-3}$ with 95\% confidence intervals. In Fig.~\ref{SEMs}f the full g-factor isosurface is shown (after subtracting a radius of 1.9 to visualize the variation). The solid lines in Figs.~\ref{SEMs}c,e correspond to $g$-factors reconstructed from the tensor. We note that $g_{xz}$ significantly differs from zero and would correspond to a 0.6~nm dipole matrix element in Ref.~\cite{Yang2013}. From the tensor we extract the Dresselhaus term associated with the vector potential $\beta_g^* = |\mu_B g^1_{xy}/2| = 109.9\pm6.9$~MHz/T resulting in $\beta_g = (178\pm 11)\times 10^{-13}$~eVcm for electrons occupying the lower energy valley. We note that due to the observed $\sin 2\varphi_B$ dependence on the $g$-factor, it is unlikely there is an interface step in the vicinity of the dot since such a step would break this periodicity~\cite{Veldhorst2015b, Ruskov2018, Ferdous2018}.

To extract the Rashba interaction strength we additionally measure the Stark shift due to the top gate voltage as a function of magnetic field in-plane angle in the same charge configuration (Fig.~\ref{SEMs}d). Subsequently we can extract the ratio between Rashba and Dresselhaus coefficients since Rashba corresponds to the offset of the sinusoidal fit $A\sin2\varphi_B+B$ and Dresselhaus corresponds to the amplitude of the total sine wave~\cite{Veldhorst2015b, Ruskov2018, Ferdous2018}. Such that $\frac{\alpha_g}{\beta_g} = \frac{B}{A} = 0.0852\pm0.0362$. This is consistent with previous observations of the Dresselhaus effect being stronger than Rashba effect in MOS dots~\cite{Ferdous2018_2}. The Rashba interaction strength is found to be $\alpha_g^* = 9.36\pm 3.96$~MHz/T and $\alpha_g = (15.2\pm6.4)\times 10^{-13}$~eVcm.

\section{$g$-factor difference}

In a silicon quantum dot qubit array, the SOI can vary from one dot to another which leads to a variaty of $g$-factors~\cite{Veldhorst2015b}. This variation in the SOI is caused by differences in the microscopic structure of the quantum dots, such as surface roughness and lattice imperfections~\cite{Ferdous2018}. Differences in $g$-factors allow one to individually address the qubits with a global MW field. On the other hand if the differences vanish it is possible to drive all qubits with one MW frequency, which is useful for scalable applications~\cite{Jones2018}. To measure the $g$-factor difference between neighboring dots we operate near the (1,0)-(0,1) anti-crossing and use the pulse sequences depicted in the Fig.~\ref{difference}d. First, the ESR frequency of QD1 is measured by pulsing between the control point C1 and readout/initialization. Then we shuttle the electron through the anti-crossing to C2 where we rotate the spin. After this we pulse back to C1 and read out the spin state. This will determine $f_\textrm{ESR}$ of QD1 and QD2 at the same magnetic field to yield the $g$-factor difference.

We probe the $g$-factor difference as a function of $\varphi_B$ at $\theta_B=0$ and as a function of $\theta_B$ at $\varphi_B = -45^\textrm{o},45^\textrm{o}$, with the Results shown in Figs.~\ref{difference}a,b. In the supplementary material we use the measured $g$-factor difference and $\hat{g}_1$ to estimate the $g$-tensor for the second dot. We can use $\hat{g}_1$ and $\hat{g}_2$ to determine the difference in the full $\varphi_B$-$\theta_B$ space. Fig.~\ref{difference}c shows the absolute value of the difference as an isosurface. From these measurements we extract the difference of the Rashba and Dresselhaus interactions between the dots to be $\Delta \alpha_g^* = 2.04$~MHz/T and $\Delta \beta_g^* = 10.07$~MHz/T. Notably, similar values for the SOI difference has been previously reported in an MOS double dot structure~\cite{Jock2018}.

\section{Coherence time}

The SOI also affects the Stark shift~\cite{Ferdous2018, Ruskov2018, Veldhorst2015b}, and is therefore related to the coherence time $T_2^*$. This is mainly because in the presence of charge noise the Stark shift causes the ESR frequency to fluctuate and the phase of the quantum state is lost. To minimize this decoherence, the Dresselhaus effect can be tuned to cancel the Rashba effect at a magic angle of magnetic field thereby minimizing the SOI. Since SOI is dominated by the Dresselhaus effect, this will happen close to the point where the magnetic field is aligned with [100] lattice direction~\cite{Ferdous2018, Ruskov2018}.

\begin{figure}
\centering
\includegraphics[width=0.5\textwidth]{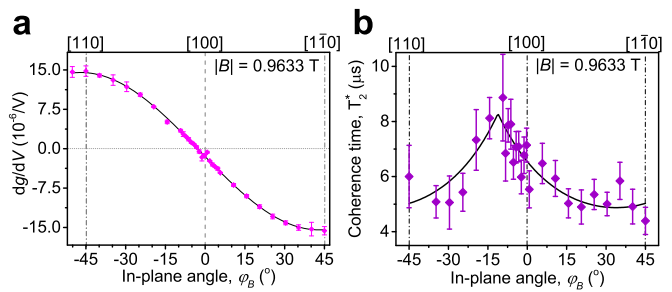}
\caption{The derivative of the Stark shift and $T^*_2$ measurement. Stark shift (a) and decoherence time $T^*_2$ (b) of the qubit defined by the G1 quantum dot in the (3,0) charge configuration as a function of the external magnetic field in-plane angle $\varphi_B$  }\label{StarkT2}
\end{figure}

In this sample, $T_2^*$ in the (1,0) charge configuration was too short to be measured reliably whereas in the (3,0) charge configuration we measure $T_2^*$ of around 5~$\mu$s~\cite{Veldhorst2015b}.  We therefore measure the Stark shift at the (3,0) charge configuration as a function of the in-plane magnetic field angle. As shown in Fig.~\ref{StarkT2}a, the Stark shift vanishes at $\varphi_B = -3^\textrm{o}$, close to the [100] lattice direction. This angle corresponds to a point where the spin orbit interaction due to the Rashba effect and Dresselhaus effect cancel. In this charge configuration, we have $\frac{\alpha_{g}}{\beta_{g}}=0.041\pm0.006$ associated with the upper energy valley state. Here, we separate two different noise sources: the decoherence caused by voltage fluctuations $\sigma_V$ and other sources. We use a simplified model for $T_2^*$ that reads
\begin{equation}
\frac{1}{T_{2}^*} = \frac{1}{T_{2\sigma_V}^*}+\frac{1}{T_{2\textrm{other}}^*},
\label{T2star}
\end{equation}
where $T_{2\sigma_V}^*$ is the coherence time that is limited by the voltage noise from the top gate and $T_{2\textrm{other}}^*$ is the coherence time limited by all the other noise sources such as magnetic noise. Now $T_{2\sigma_V}^*$ assumes the expression
\begin{equation}
T_{2\sigma_V}^* = \frac{\sqrt{2}\hbar}{\Delta F_Z |\dfrac{dg}{dF_Z}|\mu_B B},
\label{T2star}
\end{equation}
where, $\hbar$ is Planck's constant, $\Delta F_Z$ is the standard deviation of electric field along $z$-axis, $\dfrac{dg}{dF_Z}$ is the derivative of the Stark shift, and $B$ is the strength of the external magnetic field~\cite{Ferdous2018}. We assume that the only source of noise is the electrical noise along the $z$-axis. Decoherence caused by electrical noise along the $x$ and $y$ directions is significantly less prominent than along $z$-axis (see supplementary for details). 

We measure $T_2^*$ by using Ramsey interferometry (see supplementary for details) with integration times of 70 minutes. As seen in Fig.~\ref{StarkT2}b, the coherence time peaks at $\varphi_B = -10^\textrm{o}$, close to the point where the Stark shift vanishes. The difference could be caused by the fact that we are driving the transition partially electrically and this caused the Rabi frequency to vary during the measurement. The $T_2^*$ increases from around 5 $\mu$s when magnetic field is aligned along [110] up to 8.8~$\mu$s at the magic angle. From the peak point we extract $T^*_{2\textrm{,other}} = 8.2$~$\mu$s and $T^*_{2\sigma_V\textrm{,[110]}} = 15$~$\mu$s. It is worth noting that a sample where $T^*_{2\textrm{,other}}$ would be longer, the increase of $T^*_2$ could be significant~\cite{Ferdous2018}. In this device, our coherence time is 5-20 times shorter than typically measured in similar samples~\cite{Veldhorst2014,Veldhorst2015,Chan2018}. There are several possible explanations for this. Firstly, it might be due to the partially broken MW antenna which causes significant electric noise during the drive and induces additional charge noise reducing the coherence time~\cite{Muhonen2014}. Secondly, despite using isotopically enriched silicon, our decoherence might be limited by residual $^{29}$Si nuclei~\cite{Chan2018}, in which case further isotopic purification is required to reach a point where we are instead limited by the charge noise. 
\begin{figure*}
\centering
\includegraphics[width=\textwidth]{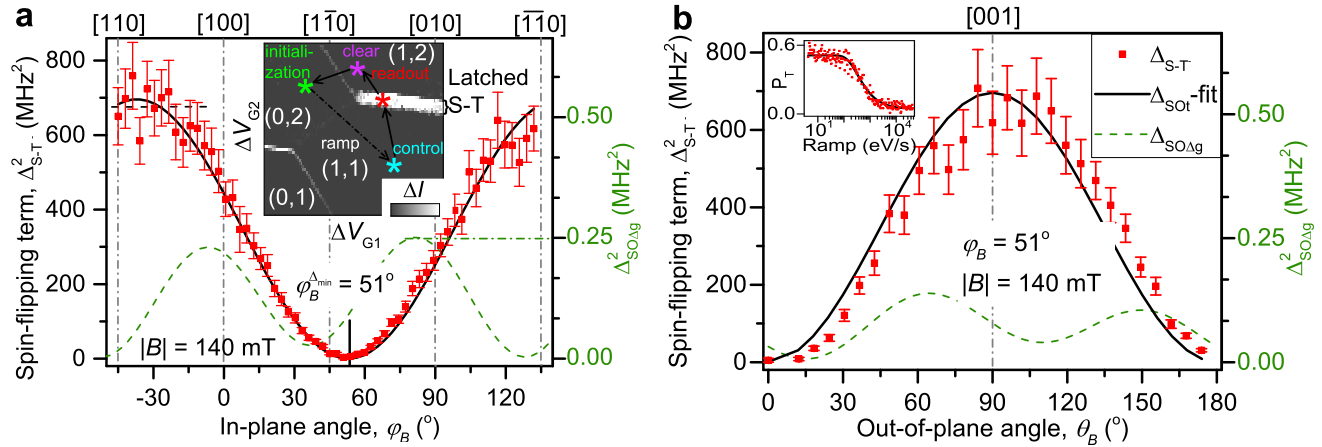}
\caption{Singlet-triplet T$^-$ mixing as a function of magnetic field direction. (a) Square of the coupling strength of singlet and triplet T$^-$ states as a function of external magnetic field in-plane angle.  Inset: Pulse sequence used to measure the coupling. (b) Square of the coupling strength of singlet and triplet T$^-$ states as a function of external magnetic field out-of-plane angle while at $\varphi_B=51^\textrm{o}$. For both (a) and (b) we show the fit based on the SOI spin-flip model and estimate from the differences in the off-diagonal terms in the $g$-tensors. Inset: Example fit of the triplet probability as a function of ramp rate across the anti-crossing.}\label{DeltaSquared}
\end{figure*}

\section{Singlet and Triplet mixing due to spin-orbit interaction}
Avoiding spin flips during tunneling is essential in proposed scalable pathways for silicon spin qubits that rely either on operating in the S-T$^0$ basis~\cite{Jones2018, Veldhorst2017} or shuttling of the spins~\cite{Taylor2005}. Two SOI mechanisms that can cause spin flipping are differences in the $g$-tensor off-diagonal terms between the two quantum dots or the induced spin-orbit field due to the interdot tunneling event. The latter spin-orbit effect has been previously studied in double GaAs dots~\cite{Hofmann2017, Scarlino2014}. Here, we consider these possible mechanisms that can cause S-T$^-$ mixing and determine their dependence on the external magnetic field direction. 

If we choose the magnetic field to be aligned with the $z$-axis, the coupling term between the S and T$^-$ states can be written as~\cite{Taylor2007, Fogarty2018}
\begin{equation}
\Delta_\textrm{S-T$^-$}(\xi) = \left|\cos(\xi)\frac{\delta E_Z^{x}+i\delta E_Z^{y}}{\sqrt{2}}+\Delta_{\textrm{SOt}}\sin(\xi)\right|,
\end{equation}
where $\delta E_Z^{n}$ is the Zeeman energy difference between the dots due to the magnetic field direction $\hat{n}$, $\Delta_{\textrm{SOt}}$ is the mixing due to the spin-orbit field caused by the movement of the electron and $\xi = -\arctan(\frac{2t_c}{E_Z})$, where $t_c$ is the tunnel coupling and $E_Z$ the Zeeman energy. The $\delta E_Z^{n}$ contributions might arise from differences in the local magnetic fields between the dots or from the differences in the corresponding terms between the $g$-tensors. Differences in the local fields could be induced by an Overhauser field due to the residual $^{29}$Si nuclei or Meissner screening~\cite{Underwood2017}. From the previous studies in isotopically enriched $^{28}$Si we would expect the contribution from a nuclear spin bath to be random in every direction with a root mean square of 50~kHz~\cite{Eng2015}. We exclude Meissner effect as a possible source of transverse fields since we operate above the critical field of the Al top gates.

In a previous study it was found that the mixing between S and T$^-$ states in the same sample was 16.4~MHz~\cite{Fogarty2018} --- significantly higher than expected from the pure hyperfine coupling due to residual $^{29}$Si nuclei in the vicinity of the dot~\cite{Eng2015}. We measure the singlet and triplet $T^-$ mixing as a function of magnetic field direction at 140~mT shown in Figs.~\ref{DeltaSquared}ab. We operate at the (1,1)-(2,0) anti-crossing with the pulsing scheme presented in the inset of Fig.~\ref{DeltaSquared}a we use electrically enhanced latched S-T readout by crossing the (1,1)-(2,1) charge transition~\cite{Fogarty2018, Harvey-Collard2018}. We initialize the singlet in the (2,0) charge state and then ramp with varying ramp rate to (1,1). With decreasing ramp rate, we observe an exponential decrease of the singlet probability seen in the inset of Fig.~\ref{DeltaSquared}b. The triplet population due to the ramp across the anti-crossing is proportional to $\exp(-2\pi|\Delta_\textrm{S-T$^-$}|/\hbar\nu)$~\cite{Shevchenko2010}, where $|\Delta_\textrm{S-T$^-$}|$ is the mixing between the S-T$^-$ states and $\nu$ is the energy level ramp rate~\cite{Fogarty2018}.

The measured dependence of the S-T$^-$ mixing term $|\Delta_\textrm{S-T$^-$}|^2$ on $\varphi_B$ and $\theta_B$ is presented in Figs.~\ref{DeltaSquared}a,b. We observe a minimal splitting of $|\Delta_\textrm{S-T$^-$,min}| = 1.85\pm 1$~MHz at $\varphi_B =51^\textrm{o}$ in-plane and  a maximum of $|\Delta_\textrm{S-T$^-$,max}| = 27.55\pm 1.6$~MHz. We exclude mixing due to the residual $^{29}$Si since we observe strong angular dependency of the mixing that is significantly higher than 50 kHz that we would expect. In Figs.~\ref{DeltaSquared}a,b we also show the expected angular dependency of S-T$^-$ mixing due to the difference in the off-diagonal elements in the $g$-tensors $\Delta_\textrm{SO$\Delta$g} = \mu_BB\frac{\delta g^{zx}+i\delta g^{zy}}{\sqrt{2}}$ (See supplementary for details). This mixing term would exhibit two local maximum and minimum in the measurement window with a maximal mixing of 500~kHz, which we do not observe. For these reasons we exclude the difference in the off-diagonal terms in $g$-tensors as a major mechanism for the mixing. In Figs.~\ref{DeltaSquared}a,b we show the fits based on a model with the spin-orbit field induced by the moving electron. One should note that here the term $\Delta_{\textrm{SOt}}$ is associated with interaction strengths $\alpha_t$ and $\beta_t$ since it arises from the $\frac{d}{dx}$ term in the Hamiltonian in Eq.~\ref{SOHamiltonian}. This fit is not unique since we have three free parameters: $\alpha_t$, $\beta_t$ and the angle of the line of dots with respect to lattice $\delta$, but only two numbers to go into the model: amplitude of the mixing and minimum point. In principle, our best guess would be based on the location of the dots based on lithography. As can be seen from Fig.~\ref{SEMs}b the angle for the line of dots with respect to lattice would correspond to $\delta = 45^\textrm{o}$, however, at this point the model diverges and we cannot give a good estimate for the values of spin-orbit coefficients. This model and divergence are discussed in more detail in the supplementary material. We conclude that the mixing is caused by the induced spin-orbit field due to the movement of the electron observable also in GaAs~\cite{Hofmann2017, Scarlino2014}. In a longer array it is possible that not all dots are aligned on the same line and hence there might be no single magnetic field direction where singlet-triplet mixing is minimized. However, appropriate top gate control might be needed to find a single magnetic field that is close to pairwise S-T mixing minimum.

\section{Conclusions}

We have studied how to control the SOI in a SiMOS double quantum dot system and use it to improve the performance of the electron spin qubits. This was achieved by probing $g$-factors, $g$-factor difference, derivative of the Stark shift, $T^*_2$ and S-T$^-$ coupling as a function of the external magnetic field direction. The Rashba and Dresselhaus interaction terms and $g$-tensors of the dots are extracted from the measurements. We also conclude that it is unlikely that there are any interface steps in the vicinity of the dots we are probing. The derivative of the Stark shift as well as the $g$-factor behaviour are in line with recent theories~\cite{Ferdous2018, Ferdous2018_2, Ruskov2018} and we observe an increase in $T_2^*$ near the direction where the derivative of the Stark shift vanishes. We also determine that S-T$^-$ mixing can be explained by the spin-orbit field due to the movement of the charge and find the angle for which this mixing is minimized. This can be used to avoid state leakage to T$^-$ when operating in the S-T$^0$ basis or in general to minimize errors when performing spin transport in a long array of dots. In addition, we could enable dynamic nuclear polarization of the residual $^{29}$Si by minimizing the mixing due to the SOI, which could otherwise quench the polarization~\cite{Nichol2015}.

As shown here, there is a trade-off between addressability and coherence time. When the magnetic field is aligned with [110] where the derivative of the Stark shift is the largest, $T_2^*$ is the shortest. Our other measurements show that the difference between the $g$-factors is the largest close to the point where the derivative of the Stark shift is the largest permitting individual addressability. The corollary is then if the external field is aligned with [100] where the derivative of the Stark shift should vanish, $T_2^*$ is longest but we lose the individual addressability. It is also possible that significantly increased $T_2^*$ (i.e. narrow ESR linewidth) could allow individual addressability if the difference between ESR frequencies is smaller~\cite{Ferdous2018}. This could be tested with a device that has a significantly longer intrinsic $T_2^*$.

It is worth noting that in a long array of dots, the $T_2^*$ in different qubits peaks at slightly different magnetic field directions depending on the individual Rashba and Dresselhaus magnitudes. Since the Rashba interaction is tunable with the top gate voltage one can align the magnetic field along [100] and tune the addressability by pulsing the gate voltages on demand. Similarly, the $g$-factor differences between qubit pairs will vanish at slightly different points. Hence, choosing an optimal field direction for a long array of dots is not trivial, but it could significantly improve the control fidelity of the qubits in the array. Understanding SOI and its impact to the qubits will be important for scaling up SiMOS qubits into a linear array, or a two-dimensional array for surface code implementation~\cite{Jones2018, Veldhorst2017, Taylor2005}. The optimal magnetic field direction for a particular choice of qubit operation mode will ultimately require careful weighing of the SOI-effects that impact upon key qubit performance parameters including gate speed, gate fidelities and state preparation and measurement fidelities. Due to the variation in the SOI and location of the dots there might not be a single 'optimal' magnetic field and that is true for all single qubits or pairwise. For a large array of dots, for instance, we could choose the average of the single or pairwise 'optimal' angle which would be significant improvement compared to the arbitrarily chosen magnetic field that is operated nowadays.


\begin{acknowledgments}

We thank W. A. Coish, A. Palyi, R. Ruskov, and C. Tahan for inspiring conversations and C. Escott for valuable feedback. We acknowledge support from the US Army Research Office (W911NF-13-1-0024 and W911NF-17-1-0198), the Australian Research Council (CE170100012), and the NSW Node of the Australian National Fabrication Facility. The views and conclusions contained in this document are those of the authors and should not be interpreted as representing the official policies, either expressed or implied, of the Army Research Office or the U.S. Government. The U.S. Government is authorized to reproduce and distribute reprints for Government purposes notwithstanding any copyright notation herein. B.H. acknowledges support from the Netherlands Organization for Scientific Research (NWO) through a Rubicon Grant. D.C. is supported by the Australian Research Council Centre for excellence in Future Low-Energy Electronics Technologies (CE170100039). K.M.I. acknowledges support from a Grant-in-Aid for Scientific Research by MEXT, NanoQuine, FIRST, and the JSPS Core-to-Core Program.

The authors declare that they have no competing financial interests.

Correspondence and requests for materials should be addressed to T.T.~(email: t.tanttu@unsw.edu.au) or A.D.~(email: a.dzurak@unsw.edu.au).
\end{acknowledgments}


\section{Appendix A}

\subsection{Device fabrication}

We fabricate our device on an epitaxially grown and isotopically enriched $^{28}$Si epilayer that has 800 ppm residue of $^{29}$Si. Four layers of gates with thicknesses 25, 60, 80, and 80 nm are fabricated on top of 5.9~nm thick SiO$_2$ with electron beam lithography and aluminium evaporation. A thermal oxide is grown on top of the Al gates to isolate layers from each other. 

\subsection{Experimental setup}

The device was bonded to a printed circuit board in a copper enclosure and cooled down in a dilution refrigerator with electron base temperature of 180~mK. The dilution refrigerator is equipped with a vector magnet that has been calibrated with Hall bars. We had small but noticeable (order of few promille) hysteresis in our superconducting coils which we observed when calibrating the magnets with Hall bars. In order to avoid the measurement this hysteresis, we always ramped in a consistent way: we started from 0 field and then ramped the field to 0.5~T to the corresponding direction and then with several steps up to 1~T. We only measure start measuring ESR frequency after 0.75~T when the hysteresis is less noticeable. Battery powered voltage sources are used to provide the DC voltages. Resistive dividers are used to combine the DC-voltage and fast gate pulses from the arbitrary waveform generator (Tektronix AWG7122C). At base temperature, the lines for S, D, G1, G2, G3 and B are filtered with a cut-off of 80 MHz and the rest of the gate lines are filtered with a cut-off of 100 Hz. A vector source (Agilent E8267D) was used to generate the MW drive. I/Q modulation of the vector source was used to introduce an adiabatic drive. MWs were attenuated by 10 dB at 4~K and 3~dB at base plate. Three different stages of adiabatic inversion drive with ranges of 10~MHz, 1~MHz, and 100~kHz were used to narrow down range for ESR frequency. SET current traces were recorded with a digital oscilloscope and analysed with a measurement computer.

Elzerman readout was used to measure the spin occupancy of the dots by pulsing between the control point and the readout/initialization point that measures the spin states and initializes spin down~\cite{Elzerman2004}. In the singlet-triplet experiment, we use latching readout in (1,2) charge occupancy, where T(1,1) are in a metastable blockade~\cite{Fogarty2018}. To determine the spin up or singlet probability, we perform 80-200 single shot readouts per point depending on the measurement.

\bibliographystyle{apsrev4-1}
\bibliography{gfactorbib_nourl}

\begin{thebibliography}{48}%
\makeatletter
\providecommand \@ifxundefined [1]{%
 \@ifx{#1\undefined}
}%
\providecommand \@ifnum [1]{%
 \ifnum #1\expandafter \@firstoftwo
 \else \expandafter \@secondoftwo
 \fi
}%
\providecommand \@ifx [1]{%
 \ifx #1\expandafter \@firstoftwo
 \else \expandafter \@secondoftwo
 \fi
}%
\providecommand \natexlab [1]{#1}%
\providecommand \enquote  [1]{``#1''}%
\providecommand \bibnamefont  [1]{#1}%
\providecommand \bibfnamefont [1]{#1}%
\providecommand \citenamefont [1]{#1}%
\providecommand \href@noop [0]{\@secondoftwo}%
\providecommand \href [0]{\begingroup \@sanitize@url \@href}%
\providecommand \@href[1]{\@@startlink{#1}\@@href}%
\providecommand \@@href[1]{\endgroup#1\@@endlink}%
\providecommand \@sanitize@url [0]{\catcode `\\12\catcode `\$12\catcode
  `\&12\catcode `\#12\catcode `\^12\catcode `\_12\catcode `\%12\relax}%
\providecommand \@@startlink[1]{}%
\providecommand \@@endlink[0]{}%
\providecommand \url  [0]{\begingroup\@sanitize@url \@url }%
\providecommand \@url [1]{\endgroup\@href {#1}{\urlprefix }}%
\providecommand \urlprefix  [0]{URL }%
\providecommand \Eprint [0]{\href }%
\providecommand \doibase [0]{http://dx.doi.org/}%
\providecommand \selectlanguage [0]{\@gobble}%
\providecommand \bibinfo  [0]{\@secondoftwo}%
\providecommand \bibfield  [0]{\@secondoftwo}%
\providecommand \translation [1]{[#1]}%
\providecommand \BibitemOpen [0]{}%
\providecommand \bibitemStop [0]{}%
\providecommand \bibitemNoStop [0]{.\EOS\space}%
\providecommand \EOS [0]{\spacefactor3000\relax}%
\providecommand \BibitemShut  [1]{\csname bibitem#1\endcsname}%
\let\auto@bib@innerbib\@empty
\bibitem [{\citenamefont {Zwanenburg}\ \emph {et~al.}(2013)\citenamefont
  {Zwanenburg}, \citenamefont {Dzurak}, \citenamefont {Morello}, \citenamefont
  {Simmons}, \citenamefont {Hollenberg}, \citenamefont {Klimeck}, \citenamefont
  {Rogge}, \citenamefont {Coppersmith},\ and\ \citenamefont
  {Eriksson}}]{Zwanenburg2013}%
  \BibitemOpen
  \bibfield  {author} {\bibinfo {author} {\bibfnamefont {F.~A.}\ \bibnamefont
  {Zwanenburg}}, \bibinfo {author} {\bibfnamefont {A.~S.}\ \bibnamefont
  {Dzurak}}, \bibinfo {author} {\bibfnamefont {A.}~\bibnamefont {Morello}},
  \bibinfo {author} {\bibfnamefont {M.~Y.}\ \bibnamefont {Simmons}}, \bibinfo
  {author} {\bibfnamefont {L.~C.~L.}\ \bibnamefont {Hollenberg}}, \bibinfo
  {author} {\bibfnamefont {G.}~\bibnamefont {Klimeck}}, \bibinfo {author}
  {\bibfnamefont {S.}~\bibnamefont {Rogge}}, \bibinfo {author} {\bibfnamefont
  {S.~N.}\ \bibnamefont {Coppersmith}}, \ and\ \bibinfo {author} {\bibfnamefont
  {M.~A.}\ \bibnamefont {Eriksson}},\ }\href@noop {} {\bibfield  {journal}
  {\bibinfo  {journal} {Rev. Mod. Phys.}\ }\textbf {\bibinfo {volume} {85}},\
  \bibinfo {pages} {961} (\bibinfo {year} {2013})}\BibitemShut {NoStop}%
\bibitem [{\citenamefont {Pla}\ \emph {et~al.}(2012)\citenamefont {Pla},
  \citenamefont {Tan}, \citenamefont {Dehollain}, \citenamefont {Lim},
  \citenamefont {Morton}, \citenamefont {Jamieson}, \citenamefont {Dzurak},\
  and\ \citenamefont {Morello}}]{Pla2012}%
  \BibitemOpen
  \bibfield  {author} {\bibinfo {author} {\bibfnamefont {J.~J.}\ \bibnamefont
  {Pla}}, \bibinfo {author} {\bibfnamefont {K.~Y.}\ \bibnamefont {Tan}},
  \bibinfo {author} {\bibfnamefont {J.~P.}\ \bibnamefont {Dehollain}}, \bibinfo
  {author} {\bibfnamefont {W.~H.}\ \bibnamefont {Lim}}, \bibinfo {author}
  {\bibfnamefont {J.~J.~L.}\ \bibnamefont {Morton}}, \bibinfo {author}
  {\bibfnamefont {D.~N.}\ \bibnamefont {Jamieson}}, \bibinfo {author}
  {\bibfnamefont {A.~S.}\ \bibnamefont {Dzurak}}, \ and\ \bibinfo {author}
  {\bibfnamefont {A.}~\bibnamefont {Morello}},\ }\href@noop {} {\bibfield
  {journal} {\bibinfo  {journal} {Nature}\ }\textbf {\bibinfo {volume} {489}},\
  \bibinfo {pages} {541} (\bibinfo {year} {2012})}\BibitemShut {NoStop}%
\bibitem [{\citenamefont {Fuechsle}\ \emph {et~al.}(2012)\citenamefont
  {Fuechsle}, \citenamefont {Miwa}, \citenamefont {Mahapatra}, \citenamefont
  {Ryu}, \citenamefont {Lee}, \citenamefont {Warschkow}, \citenamefont
  {Hollenberg}, \citenamefont {Klimeck},\ and\ \citenamefont
  {Simmons}}]{Fuechsle2012}%
  \BibitemOpen
  \bibfield  {author} {\bibinfo {author} {\bibfnamefont {M.}~\bibnamefont
  {Fuechsle}}, \bibinfo {author} {\bibfnamefont {J.~A.}\ \bibnamefont {Miwa}},
  \bibinfo {author} {\bibfnamefont {S.}~\bibnamefont {Mahapatra}}, \bibinfo
  {author} {\bibfnamefont {H.}~\bibnamefont {Ryu}}, \bibinfo {author}
  {\bibfnamefont {S.}~\bibnamefont {Lee}}, \bibinfo {author} {\bibfnamefont
  {O.}~\bibnamefont {Warschkow}}, \bibinfo {author} {\bibfnamefont {L.~C.~L.}\
  \bibnamefont {Hollenberg}}, \bibinfo {author} {\bibfnamefont
  {G.}~\bibnamefont {Klimeck}}, \ and\ \bibinfo {author} {\bibfnamefont
  {M.~Y.}\ \bibnamefont {Simmons}},\ }\href@noop {} {\bibfield  {journal}
  {\bibinfo  {journal} {Nat. Nanotechnol.}\ }\textbf {\bibinfo {volume} {7}},\
  \bibinfo {pages} {242} (\bibinfo {year} {2012})}\BibitemShut {NoStop}%
\bibitem [{\citenamefont {Veldhorst}\ \emph
  {et~al.}(2015{\natexlab{a}})\citenamefont {Veldhorst}, \citenamefont {Yang},
  \citenamefont {Hwang}, \citenamefont {Huang}, \citenamefont {Dehollain},
  \citenamefont {Muhonen}, \citenamefont {Simmons}, \citenamefont {Laucht},
  \citenamefont {Hudson}, \citenamefont {Itoh}, \citenamefont {Morello},\ and\
  \citenamefont {Dzurak}}]{Veldhorst2015}%
  \BibitemOpen
  \bibfield  {author} {\bibinfo {author} {\bibfnamefont {M.}~\bibnamefont
  {Veldhorst}}, \bibinfo {author} {\bibfnamefont {C.~H.}\ \bibnamefont {Yang}},
  \bibinfo {author} {\bibfnamefont {J.~C.~C.}\ \bibnamefont {Hwang}}, \bibinfo
  {author} {\bibfnamefont {W.}~\bibnamefont {Huang}}, \bibinfo {author}
  {\bibfnamefont {J.~P.}\ \bibnamefont {Dehollain}}, \bibinfo {author}
  {\bibfnamefont {J.~T.}\ \bibnamefont {Muhonen}}, \bibinfo {author}
  {\bibfnamefont {S.}~\bibnamefont {Simmons}}, \bibinfo {author} {\bibfnamefont
  {A.}~\bibnamefont {Laucht}}, \bibinfo {author} {\bibfnamefont {F.~E.}\
  \bibnamefont {Hudson}}, \bibinfo {author} {\bibfnamefont {K.~M.}\
  \bibnamefont {Itoh}}, \bibinfo {author} {\bibfnamefont {A.}~\bibnamefont
  {Morello}}, \ and\ \bibinfo {author} {\bibfnamefont {A.~S.}\ \bibnamefont
  {Dzurak}},\ }\href@noop {} {\bibfield  {journal} {\bibinfo  {journal}
  {Nature}\ }\textbf {\bibinfo {volume} {526}},\ \bibinfo {pages} {410}
  (\bibinfo {year} {2015}{\natexlab{a}})}\BibitemShut {NoStop}%
\bibitem [{\citenamefont {Bienfait}\ \emph {et~al.}(2015)\citenamefont
  {Bienfait}, \citenamefont {Pla}, \citenamefont {Kubo}, \citenamefont {Stern},
  \citenamefont {Zhou}, \citenamefont {Lo}, \citenamefont {Weis}, \citenamefont
  {Schenkel}, \citenamefont {Thewalt}, \citenamefont {Vion}, \citenamefont
  {Esteve}, \citenamefont {Julsgaard}, \citenamefont {M{\o}lmer}, \citenamefont
  {Morton},\ and\ \citenamefont {Bertet}}]{Bienfait2015}%
  \BibitemOpen
  \bibfield  {author} {\bibinfo {author} {\bibfnamefont {A.}~\bibnamefont
  {Bienfait}}, \bibinfo {author} {\bibfnamefont {J.~J.}\ \bibnamefont {Pla}},
  \bibinfo {author} {\bibfnamefont {Y.}~\bibnamefont {Kubo}}, \bibinfo {author}
  {\bibfnamefont {M.}~\bibnamefont {Stern}}, \bibinfo {author} {\bibfnamefont
  {X.}~\bibnamefont {Zhou}}, \bibinfo {author} {\bibfnamefont {C.~C.}\
  \bibnamefont {Lo}}, \bibinfo {author} {\bibfnamefont {C.~D.}\ \bibnamefont
  {Weis}}, \bibinfo {author} {\bibfnamefont {T.}~\bibnamefont {Schenkel}},
  \bibinfo {author} {\bibfnamefont {M.~L.~W.}\ \bibnamefont {Thewalt}},
  \bibinfo {author} {\bibfnamefont {D.}~\bibnamefont {Vion}}, \bibinfo {author}
  {\bibfnamefont {D.}~\bibnamefont {Esteve}}, \bibinfo {author} {\bibfnamefont
  {B.}~\bibnamefont {Julsgaard}}, \bibinfo {author} {\bibfnamefont
  {K.}~\bibnamefont {M{\o}lmer}}, \bibinfo {author} {\bibfnamefont {J.~J.~L.}\
  \bibnamefont {Morton}}, \ and\ \bibinfo {author} {\bibfnamefont
  {P.}~\bibnamefont {Bertet}},\ }\href@noop {} {\bibfield  {journal} {\bibinfo
  {journal} {Nat. Nanotechnol.}\ }\textbf {\bibinfo {volume} {11}},\ \bibinfo
  {pages} {253} (\bibinfo {year} {2015})}\BibitemShut {NoStop}%
\bibitem [{\citenamefont {Mi}\ \emph {et~al.}(2016)\citenamefont {Mi},
  \citenamefont {Cady}, \citenamefont {Zajac}, \citenamefont {Deelman},\ and\
  \citenamefont {Petta}}]{Mi2016}%
  \BibitemOpen
  \bibfield  {author} {\bibinfo {author} {\bibfnamefont {X.}~\bibnamefont
  {Mi}}, \bibinfo {author} {\bibfnamefont {J.~V.}\ \bibnamefont {Cady}},
  \bibinfo {author} {\bibfnamefont {D.~M.}\ \bibnamefont {Zajac}}, \bibinfo
  {author} {\bibfnamefont {P.~W.}\ \bibnamefont {Deelman}}, \ and\ \bibinfo
  {author} {\bibfnamefont {J.~R.}\ \bibnamefont {Petta}},\ }\href@noop {}
  {\bibfield  {journal} {\bibinfo  {journal} {Science}\ }\textbf {\bibinfo
  {volume} {355}} (\bibinfo {year} {2016})}\BibitemShut {NoStop}%
\bibitem [{\citenamefont {Maurand}\ \emph {et~al.}(2016)\citenamefont
  {Maurand}, \citenamefont {Jehl}, \citenamefont {Kotekar-Patil}, \citenamefont
  {Corna}, \citenamefont {Bohuslavskyi}, \citenamefont {Lavi{\'e}ville},
  \citenamefont {Hutin}, \citenamefont {Barraud}, \citenamefont {Vinet},
  \citenamefont {Sanquer},\ and\ \citenamefont {De~Franceschi}}]{Maurand2016}%
  \BibitemOpen
  \bibfield  {author} {\bibinfo {author} {\bibfnamefont {R.}~\bibnamefont
  {Maurand}}, \bibinfo {author} {\bibfnamefont {X.}~\bibnamefont {Jehl}},
  \bibinfo {author} {\bibfnamefont {D.}~\bibnamefont {Kotekar-Patil}}, \bibinfo
  {author} {\bibfnamefont {A.}~\bibnamefont {Corna}}, \bibinfo {author}
  {\bibfnamefont {H.}~\bibnamefont {Bohuslavskyi}}, \bibinfo {author}
  {\bibfnamefont {R.}~\bibnamefont {Lavi{\'e}ville}}, \bibinfo {author}
  {\bibfnamefont {L.}~\bibnamefont {Hutin}}, \bibinfo {author} {\bibfnamefont
  {S.}~\bibnamefont {Barraud}}, \bibinfo {author} {\bibfnamefont
  {M.}~\bibnamefont {Vinet}}, \bibinfo {author} {\bibfnamefont
  {M.}~\bibnamefont {Sanquer}}, \ and\ \bibinfo {author} {\bibfnamefont
  {S.}~\bibnamefont {De~Franceschi}},\ }\href@noop {} {\bibfield  {journal}
  {\bibinfo  {journal} {Nat. Commun.}\ }\textbf {\bibinfo {volume} {7}},\
  \bibinfo {pages} {13575} (\bibinfo {year} {2016})},\ \bibinfo {note}
  {article}\BibitemShut {NoStop}%
\bibitem [{\citenamefont {Fujita}\ \emph {et~al.}(2017)\citenamefont {Fujita},
  \citenamefont {Baart}, \citenamefont {Reichl}, \citenamefont {Wegscheider},\
  and\ \citenamefont {Vandersypen}}]{Fujita2017}%
  \BibitemOpen
  \bibfield  {author} {\bibinfo {author} {\bibfnamefont {T.}~\bibnamefont
  {Fujita}}, \bibinfo {author} {\bibfnamefont {T.~A.}\ \bibnamefont {Baart}},
  \bibinfo {author} {\bibfnamefont {C.}~\bibnamefont {Reichl}}, \bibinfo
  {author} {\bibfnamefont {W.}~\bibnamefont {Wegscheider}}, \ and\ \bibinfo
  {author} {\bibfnamefont {L.~M.~K.}\ \bibnamefont {Vandersypen}},\ }\href@noop
  {} {\bibfield  {journal} {\bibinfo  {journal} {npj Quantum Information}\
  }\textbf {\bibinfo {volume} {3}},\ \bibinfo {pages} {22} (\bibinfo {year}
  {2017})}\BibitemShut {NoStop}%
\bibitem [{\citenamefont {Samkharadze}\ \emph {et~al.}(2018)\citenamefont
  {Samkharadze}, \citenamefont {Zheng}, \citenamefont {Kalhor}, \citenamefont
  {Brousse}, \citenamefont {Sammak}, \citenamefont {Mendes}, \citenamefont
  {Blais}, \citenamefont {Scappucci},\ and\ \citenamefont
  {Vandersypen}}]{Samkharadze2018}%
  \BibitemOpen
  \bibfield  {author} {\bibinfo {author} {\bibfnamefont {N.}~\bibnamefont
  {Samkharadze}}, \bibinfo {author} {\bibfnamefont {G.}~\bibnamefont {Zheng}},
  \bibinfo {author} {\bibfnamefont {N.}~\bibnamefont {Kalhor}}, \bibinfo
  {author} {\bibfnamefont {D.}~\bibnamefont {Brousse}}, \bibinfo {author}
  {\bibfnamefont {A.}~\bibnamefont {Sammak}}, \bibinfo {author} {\bibfnamefont
  {U.~C.}\ \bibnamefont {Mendes}}, \bibinfo {author} {\bibfnamefont
  {A.}~\bibnamefont {Blais}}, \bibinfo {author} {\bibfnamefont
  {G.}~\bibnamefont {Scappucci}}, \ and\ \bibinfo {author} {\bibfnamefont
  {L.~M.~K.}\ \bibnamefont {Vandersypen}},\ }\href@noop {} {\bibfield
  {journal} {\bibinfo  {journal} {Science}\ }\textbf {\bibinfo {volume} {359}}
  (\bibinfo {year} {2018})}\BibitemShut {NoStop}%
\bibitem [{\citenamefont {Watson}\ \emph {et~al.}(2018)\citenamefont {Watson},
  \citenamefont {Philips}, \citenamefont {Kawakami}, \citenamefont {Ward},
  \citenamefont {Scarlino}, \citenamefont {Veldhorst}, \citenamefont {Savage},
  \citenamefont {Lagally}, \citenamefont {Friesen}, \citenamefont
  {Coppersmith}, \citenamefont {Eriksson},\ and\ \citenamefont
  {Vandersypen}}]{Watson2018}%
  \BibitemOpen
  \bibfield  {author} {\bibinfo {author} {\bibfnamefont {T.~F.}\ \bibnamefont
  {Watson}}, \bibinfo {author} {\bibfnamefont {S.~G.~J.}\ \bibnamefont
  {Philips}}, \bibinfo {author} {\bibfnamefont {E.}~\bibnamefont {Kawakami}},
  \bibinfo {author} {\bibfnamefont {D.~R.}\ \bibnamefont {Ward}}, \bibinfo
  {author} {\bibfnamefont {P.}~\bibnamefont {Scarlino}}, \bibinfo {author}
  {\bibfnamefont {M.}~\bibnamefont {Veldhorst}}, \bibinfo {author}
  {\bibfnamefont {D.~E.}\ \bibnamefont {Savage}}, \bibinfo {author}
  {\bibfnamefont {M.~G.}\ \bibnamefont {Lagally}}, \bibinfo {author}
  {\bibfnamefont {M.}~\bibnamefont {Friesen}}, \bibinfo {author} {\bibfnamefont
  {S.~N.}\ \bibnamefont {Coppersmith}}, \bibinfo {author} {\bibfnamefont
  {M.~A.}\ \bibnamefont {Eriksson}}, \ and\ \bibinfo {author} {\bibfnamefont
  {L.~M.~K.}\ \bibnamefont {Vandersypen}},\ }\href@noop {} {\bibfield
  {journal} {\bibinfo  {journal} {Nature}\ } (\bibinfo {year}
  {2018})}\BibitemShut {NoStop}%
\bibitem [{\citenamefont {Zajac}\ \emph {et~al.}(2018)\citenamefont {Zajac},
  \citenamefont {Sigillito}, \citenamefont {Russ}, \citenamefont {Borjans},
  \citenamefont {Taylor}, \citenamefont {Burkard},\ and\ \citenamefont
  {Petta}}]{Zajac2018}%
  \BibitemOpen
  \bibfield  {author} {\bibinfo {author} {\bibfnamefont {D.~M.}\ \bibnamefont
  {Zajac}}, \bibinfo {author} {\bibfnamefont {A.~J.}\ \bibnamefont
  {Sigillito}}, \bibinfo {author} {\bibfnamefont {M.}~\bibnamefont {Russ}},
  \bibinfo {author} {\bibfnamefont {F.}~\bibnamefont {Borjans}}, \bibinfo
  {author} {\bibfnamefont {J.~M.}\ \bibnamefont {Taylor}}, \bibinfo {author}
  {\bibfnamefont {G.}~\bibnamefont {Burkard}}, \ and\ \bibinfo {author}
  {\bibfnamefont {J.~R.}\ \bibnamefont {Petta}},\ }\href@noop {} {\bibfield
  {journal} {\bibinfo  {journal} {Science}\ }\textbf {\bibinfo {volume}
  {359}},\ \bibinfo {pages} {439} (\bibinfo {year} {2018})}\BibitemShut
  {NoStop}%
\bibitem [{\citenamefont {Jock}\ \emph {et~al.}(2018)\citenamefont {Jock},
  \citenamefont {Jacobson}, \citenamefont {Harvey-Collard}, \citenamefont
  {Mounce}, \citenamefont {Srinivasa}, \citenamefont {Ward}, \citenamefont
  {Anderson}, \citenamefont {Manginell}, \citenamefont {Wendt}, \citenamefont
  {Rudolph}, \citenamefont {Pluym}, \citenamefont {Gamble}, \citenamefont
  {Baczewski}, \citenamefont {Witzel},\ and\ \citenamefont
  {Carroll}}]{Jock2018}%
  \BibitemOpen
  \bibfield  {author} {\bibinfo {author} {\bibfnamefont {R.~M.}\ \bibnamefont
  {Jock}}, \bibinfo {author} {\bibfnamefont {N.~T.}\ \bibnamefont {Jacobson}},
  \bibinfo {author} {\bibfnamefont {P.}~\bibnamefont {Harvey-Collard}},
  \bibinfo {author} {\bibfnamefont {A.~M.}\ \bibnamefont {Mounce}}, \bibinfo
  {author} {\bibfnamefont {V.}~\bibnamefont {Srinivasa}}, \bibinfo {author}
  {\bibfnamefont {D.~R.}\ \bibnamefont {Ward}}, \bibinfo {author}
  {\bibfnamefont {J.}~\bibnamefont {Anderson}}, \bibinfo {author}
  {\bibfnamefont {R.}~\bibnamefont {Manginell}}, \bibinfo {author}
  {\bibfnamefont {J.~R.}\ \bibnamefont {Wendt}}, \bibinfo {author}
  {\bibfnamefont {M.}~\bibnamefont {Rudolph}}, \bibinfo {author} {\bibfnamefont
  {T.}~\bibnamefont {Pluym}}, \bibinfo {author} {\bibfnamefont {J.~K.}\
  \bibnamefont {Gamble}}, \bibinfo {author} {\bibfnamefont {A.~D.}\
  \bibnamefont {Baczewski}}, \bibinfo {author} {\bibfnamefont {W.~M.}\
  \bibnamefont {Witzel}}, \ and\ \bibinfo {author} {\bibfnamefont {M.~S.}\
  \bibnamefont {Carroll}},\ }\href {\doibase 10.1038/s41467-018-04200-0}
  {\bibfield  {journal} {\bibinfo  {journal} {Nat. Commun.}\ }\textbf {\bibinfo
  {volume} {9}},\ \bibinfo {pages} {1768} (\bibinfo {year} {2018})}\BibitemShut
  {NoStop}%
\bibitem [{\citenamefont {Nakajima}\ \emph {et~al.}(2018)\citenamefont
  {Nakajima}, \citenamefont {Delbecq}, \citenamefont {Otsuka}, \citenamefont
  {Amaha}, \citenamefont {Yoneda}, \citenamefont {Noiri}, \citenamefont
  {Takeda}, \citenamefont {Allison}, \citenamefont {Ludwig}, \citenamefont
  {Wieck}, \citenamefont {Hu}, \citenamefont {Nori},\ and\ \citenamefont
  {Tarucha}}]{Nakajima2018}%
  \BibitemOpen
  \bibfield  {author} {\bibinfo {author} {\bibfnamefont {T.}~\bibnamefont
  {Nakajima}}, \bibinfo {author} {\bibfnamefont {M.~R.}\ \bibnamefont
  {Delbecq}}, \bibinfo {author} {\bibfnamefont {T.}~\bibnamefont {Otsuka}},
  \bibinfo {author} {\bibfnamefont {S.}~\bibnamefont {Amaha}}, \bibinfo
  {author} {\bibfnamefont {J.}~\bibnamefont {Yoneda}}, \bibinfo {author}
  {\bibfnamefont {A.}~\bibnamefont {Noiri}}, \bibinfo {author} {\bibfnamefont
  {K.}~\bibnamefont {Takeda}}, \bibinfo {author} {\bibfnamefont
  {G.}~\bibnamefont {Allison}}, \bibinfo {author} {\bibfnamefont
  {A.}~\bibnamefont {Ludwig}}, \bibinfo {author} {\bibfnamefont {A.~D.}\
  \bibnamefont {Wieck}}, \bibinfo {author} {\bibfnamefont {X.}~\bibnamefont
  {Hu}}, \bibinfo {author} {\bibfnamefont {F.}~\bibnamefont {Nori}}, \ and\
  \bibinfo {author} {\bibfnamefont {S.}~\bibnamefont {Tarucha}},\ }\href@noop
  {} {\bibfield  {journal} {\bibinfo  {journal} {Nat. Commun.}\ }\textbf
  {\bibinfo {volume} {9}},\ \bibinfo {pages} {2133} (\bibinfo {year}
  {2018})}\BibitemShut {NoStop}%
\bibitem [{\citenamefont {Veldhorst}\ \emph
  {et~al.}(2015{\natexlab{b}})\citenamefont {Veldhorst}, \citenamefont
  {Ruskov}, \citenamefont {Yang}, \citenamefont {Hwang}, \citenamefont
  {Hudson}, \citenamefont {Flatt\'e}, \citenamefont {Tahan}, \citenamefont
  {Itoh}, \citenamefont {Morello},\ and\ \citenamefont
  {Dzurak}}]{Veldhorst2015b}%
  \BibitemOpen
  \bibfield  {author} {\bibinfo {author} {\bibfnamefont {M.}~\bibnamefont
  {Veldhorst}}, \bibinfo {author} {\bibfnamefont {R.}~\bibnamefont {Ruskov}},
  \bibinfo {author} {\bibfnamefont {C.~H.}\ \bibnamefont {Yang}}, \bibinfo
  {author} {\bibfnamefont {J.~C.~C.}\ \bibnamefont {Hwang}}, \bibinfo {author}
  {\bibfnamefont {F.~E.}\ \bibnamefont {Hudson}}, \bibinfo {author}
  {\bibfnamefont {M.~E.}\ \bibnamefont {Flatt\'e}}, \bibinfo {author}
  {\bibfnamefont {C.}~\bibnamefont {Tahan}}, \bibinfo {author} {\bibfnamefont
  {K.~M.}\ \bibnamefont {Itoh}}, \bibinfo {author} {\bibfnamefont
  {A.}~\bibnamefont {Morello}}, \ and\ \bibinfo {author} {\bibfnamefont
  {A.~S.}\ \bibnamefont {Dzurak}},\ }\href@noop {} {\bibfield  {journal}
  {\bibinfo  {journal} {Phys. Rev. B}\ }\textbf {\bibinfo {volume} {92}},\
  \bibinfo {pages} {201401} (\bibinfo {year} {2015}{\natexlab{b}})}\BibitemShut
  {NoStop}%
\bibitem [{\citenamefont {Ruskov}\ \emph {et~al.}(2018)\citenamefont {Ruskov},
  \citenamefont {Veldhorst}, \citenamefont {Dzurak},\ and\ \citenamefont
  {Tahan}}]{Ruskov2018}%
  \BibitemOpen
  \bibfield  {author} {\bibinfo {author} {\bibfnamefont {R.}~\bibnamefont
  {Ruskov}}, \bibinfo {author} {\bibfnamefont {M.}~\bibnamefont {Veldhorst}},
  \bibinfo {author} {\bibfnamefont {A.~S.}\ \bibnamefont {Dzurak}}, \ and\
  \bibinfo {author} {\bibfnamefont {C.}~\bibnamefont {Tahan}},\ }\href@noop {}
  {\bibfield  {journal} {\bibinfo  {journal} {Phys. Rev. B}\ }\textbf {\bibinfo
  {volume} {98}},\ \bibinfo {pages} {245424} (\bibinfo {year}
  {2018})}\BibitemShut {NoStop}%
\bibitem [{\citenamefont {Ferdous}\ \emph
  {et~al.}(2018{\natexlab{a}})\citenamefont {Ferdous}, \citenamefont {Chan},
  \citenamefont {Veldhorst}, \citenamefont {Hwang}, \citenamefont {Yang},
  \citenamefont {Sahasrabudhe}, \citenamefont {Klimeck}, \citenamefont
  {Morello}, \citenamefont {Dzurak},\ and\ \citenamefont
  {Rahman}}]{Ferdous2018}%
  \BibitemOpen
  \bibfield  {author} {\bibinfo {author} {\bibfnamefont {R.}~\bibnamefont
  {Ferdous}}, \bibinfo {author} {\bibfnamefont {K.~W.}\ \bibnamefont {Chan}},
  \bibinfo {author} {\bibfnamefont {M.}~\bibnamefont {Veldhorst}}, \bibinfo
  {author} {\bibfnamefont {J.~C.~C.}\ \bibnamefont {Hwang}}, \bibinfo {author}
  {\bibfnamefont {C.~H.}\ \bibnamefont {Yang}}, \bibinfo {author}
  {\bibfnamefont {H.}~\bibnamefont {Sahasrabudhe}}, \bibinfo {author}
  {\bibfnamefont {G.}~\bibnamefont {Klimeck}}, \bibinfo {author} {\bibfnamefont
  {A.}~\bibnamefont {Morello}}, \bibinfo {author} {\bibfnamefont {A.~S.}\
  \bibnamefont {Dzurak}}, \ and\ \bibinfo {author} {\bibfnamefont
  {R.}~\bibnamefont {Rahman}},\ }\href@noop {} {\bibfield  {journal} {\bibinfo
  {journal} {Phys. Rev. B}\ }\textbf {\bibinfo {volume} {97}},\ \bibinfo
  {pages} {241401} (\bibinfo {year} {2018}{\natexlab{a}})}\BibitemShut
  {NoStop}%
\bibitem [{\citenamefont {Ferdous}\ \emph
  {et~al.}(2018{\natexlab{b}})\citenamefont {Ferdous}, \citenamefont
  {Kawakami}, \citenamefont {Scarlino}, \citenamefont {Nowak}, \citenamefont
  {Ward}, \citenamefont {Savage}, \citenamefont {Lagally}, \citenamefont
  {Coppersmith}, \citenamefont {Friesen}, \citenamefont {Eriksson},
  \citenamefont {Vandersypen},\ and\ \citenamefont {Rahman}}]{Ferdous2018_2}%
  \BibitemOpen
  \bibfield  {author} {\bibinfo {author} {\bibfnamefont {R.}~\bibnamefont
  {Ferdous}}, \bibinfo {author} {\bibfnamefont {E.}~\bibnamefont {Kawakami}},
  \bibinfo {author} {\bibfnamefont {P.}~\bibnamefont {Scarlino}}, \bibinfo
  {author} {\bibfnamefont {M.~P.}\ \bibnamefont {Nowak}}, \bibinfo {author}
  {\bibfnamefont {D.~R.}\ \bibnamefont {Ward}}, \bibinfo {author}
  {\bibfnamefont {D.~E.}\ \bibnamefont {Savage}}, \bibinfo {author}
  {\bibfnamefont {M.~G.}\ \bibnamefont {Lagally}}, \bibinfo {author}
  {\bibfnamefont {S.~N.}\ \bibnamefont {Coppersmith}}, \bibinfo {author}
  {\bibfnamefont {M.}~\bibnamefont {Friesen}}, \bibinfo {author} {\bibfnamefont
  {M.~A.}\ \bibnamefont {Eriksson}}, \bibinfo {author} {\bibfnamefont
  {L.~M.~K.}\ \bibnamefont {Vandersypen}}, \ and\ \bibinfo {author}
  {\bibfnamefont {R.}~\bibnamefont {Rahman}},\ }\href@noop {} {\bibfield
  {journal} {\bibinfo  {journal} {npj Quantum Information}\ }\textbf {\bibinfo
  {volume} {4}},\ \bibinfo {pages} {26} (\bibinfo {year}
  {2018}{\natexlab{b}})}\BibitemShut {NoStop}%
\bibitem [{\citenamefont {Stepanenko}\ \emph {et~al.}(2012)\citenamefont
  {Stepanenko}, \citenamefont {Rudner}, \citenamefont {Halperin},\ and\
  \citenamefont {Loss}}]{Stepanenko2012}%
  \BibitemOpen
  \bibfield  {author} {\bibinfo {author} {\bibfnamefont {D.}~\bibnamefont
  {Stepanenko}}, \bibinfo {author} {\bibfnamefont {M.}~\bibnamefont {Rudner}},
  \bibinfo {author} {\bibfnamefont {B.~I.}\ \bibnamefont {Halperin}}, \ and\
  \bibinfo {author} {\bibfnamefont {D.}~\bibnamefont {Loss}},\ }\href@noop {}
  {\bibfield  {journal} {\bibinfo  {journal} {Phys. Rev. B}\ }\textbf {\bibinfo
  {volume} {85}},\ \bibinfo {pages} {075416} (\bibinfo {year}
  {2012})}\BibitemShut {NoStop}%
\bibitem [{\citenamefont {Corna}\ \emph {et~al.}(2018)\citenamefont {Corna},
  \citenamefont {Bourdet}, \citenamefont {Maurand}, \citenamefont {Crippa},
  \citenamefont {Kotekar-Patil}, \citenamefont {Bohuslavskyi}, \citenamefont
  {Lavi{\'e}ville}, \citenamefont {Hutin}, \citenamefont {Barraud},
  \citenamefont {Jehl}, \citenamefont {Vinet}, \citenamefont {De~Franceschi},
  \citenamefont {Niquet},\ and\ \citenamefont {Sanquer}}]{Corna2018}%
  \BibitemOpen
  \bibfield  {author} {\bibinfo {author} {\bibfnamefont {A.}~\bibnamefont
  {Corna}}, \bibinfo {author} {\bibfnamefont {L.}~\bibnamefont {Bourdet}},
  \bibinfo {author} {\bibfnamefont {R.}~\bibnamefont {Maurand}}, \bibinfo
  {author} {\bibfnamefont {A.}~\bibnamefont {Crippa}}, \bibinfo {author}
  {\bibfnamefont {D.}~\bibnamefont {Kotekar-Patil}}, \bibinfo {author}
  {\bibfnamefont {H.}~\bibnamefont {Bohuslavskyi}}, \bibinfo {author}
  {\bibfnamefont {R.}~\bibnamefont {Lavi{\'e}ville}}, \bibinfo {author}
  {\bibfnamefont {L.}~\bibnamefont {Hutin}}, \bibinfo {author} {\bibfnamefont
  {S.}~\bibnamefont {Barraud}}, \bibinfo {author} {\bibfnamefont
  {X.}~\bibnamefont {Jehl}}, \bibinfo {author} {\bibfnamefont {M.}~\bibnamefont
  {Vinet}}, \bibinfo {author} {\bibfnamefont {S.}~\bibnamefont
  {De~Franceschi}}, \bibinfo {author} {\bibfnamefont {Y.-M.}\ \bibnamefont
  {Niquet}}, \ and\ \bibinfo {author} {\bibfnamefont {M.}~\bibnamefont
  {Sanquer}},\ }\href@noop {} {\bibfield  {journal} {\bibinfo  {journal} {npj
  Quantum Information}\ }\textbf {\bibinfo {volume} {4}},\ \bibinfo {pages} {6}
  (\bibinfo {year} {2018})}\BibitemShut {NoStop}%
\bibitem [{\citenamefont {Huang}\ \emph {et~al.}(2017)\citenamefont {Huang},
  \citenamefont {Veldhorst}, \citenamefont {Zimmerman}, \citenamefont
  {Dzurak},\ and\ \citenamefont {Culcer}}]{Huang2017}%
  \BibitemOpen
  \bibfield  {author} {\bibinfo {author} {\bibfnamefont {W.}~\bibnamefont
  {Huang}}, \bibinfo {author} {\bibfnamefont {M.}~\bibnamefont {Veldhorst}},
  \bibinfo {author} {\bibfnamefont {N.~M.}\ \bibnamefont {Zimmerman}}, \bibinfo
  {author} {\bibfnamefont {A.~S.}\ \bibnamefont {Dzurak}}, \ and\ \bibinfo
  {author} {\bibfnamefont {D.}~\bibnamefont {Culcer}},\ }\href@noop {}
  {\bibfield  {journal} {\bibinfo  {journal} {Phys. Rev. B}\ }\textbf {\bibinfo
  {volume} {95}},\ \bibinfo {pages} {075403} (\bibinfo {year}
  {2017})}\BibitemShut {NoStop}%
\bibitem [{\citenamefont {Huang}\ \emph {et~al.}(2018)\citenamefont {Huang},
  \citenamefont {Yang}, \citenamefont {Chan}, \citenamefont {Tanttu},
  \citenamefont {Hensen}, \citenamefont {Leon}, \citenamefont {Fogarty},
  \citenamefont {Hwang}, \citenamefont {Hudson}, \citenamefont {Itoh},
  \citenamefont {Morello}, \citenamefont {Laucht},\ and\ \citenamefont
  {Dzurak}}]{Huang2018arxiv}%
  \BibitemOpen
  \bibfield  {author} {\bibinfo {author} {\bibfnamefont {W.}~\bibnamefont
  {Huang}}, \bibinfo {author} {\bibfnamefont {C.~H.}\ \bibnamefont {Yang}},
  \bibinfo {author} {\bibfnamefont {K.~W.}\ \bibnamefont {Chan}}, \bibinfo
  {author} {\bibfnamefont {T.}~\bibnamefont {Tanttu}}, \bibinfo {author}
  {\bibfnamefont {B.}~\bibnamefont {Hensen}}, \bibinfo {author} {\bibfnamefont
  {R.~C.~C.}\ \bibnamefont {Leon}}, \bibinfo {author} {\bibfnamefont {M.~A.}\
  \bibnamefont {Fogarty}}, \bibinfo {author} {\bibfnamefont {J.~C.~C.}\
  \bibnamefont {Hwang}}, \bibinfo {author} {\bibfnamefont {F.~E.}\ \bibnamefont
  {Hudson}}, \bibinfo {author} {\bibfnamefont {K.~M.}\ \bibnamefont {Itoh}},
  \bibinfo {author} {\bibfnamefont {A.}~\bibnamefont {Morello}}, \bibinfo
  {author} {\bibfnamefont {A.}~\bibnamefont {Laucht}}, \ and\ \bibinfo {author}
  {\bibfnamefont {A.~S.}\ \bibnamefont {Dzurak}},\ }\href@noop {} {\bibfield
  {journal} {\bibinfo  {journal} {arXiv:}\ }\textbf {\bibinfo {volume}
  {1805.05027}} (\bibinfo {year} {2018})}\BibitemShut {NoStop}%
\bibitem [{\citenamefont {Hofmann}\ \emph {et~al.}(2017)\citenamefont
  {Hofmann}, \citenamefont {Maisi}, \citenamefont {Kr\"ahenmann}, \citenamefont
  {Reichl}, \citenamefont {Wegscheider}, \citenamefont {Ensslin},\ and\
  \citenamefont {Ihn}}]{Hofmann2017}%
  \BibitemOpen
  \bibfield  {author} {\bibinfo {author} {\bibfnamefont {A.}~\bibnamefont
  {Hofmann}}, \bibinfo {author} {\bibfnamefont {V.~F.}\ \bibnamefont {Maisi}},
  \bibinfo {author} {\bibfnamefont {T.}~\bibnamefont {Kr\"ahenmann}}, \bibinfo
  {author} {\bibfnamefont {C.}~\bibnamefont {Reichl}}, \bibinfo {author}
  {\bibfnamefont {W.}~\bibnamefont {Wegscheider}}, \bibinfo {author}
  {\bibfnamefont {K.}~\bibnamefont {Ensslin}}, \ and\ \bibinfo {author}
  {\bibfnamefont {T.}~\bibnamefont {Ihn}},\ }\href@noop {} {\bibfield
  {journal} {\bibinfo  {journal} {Phys. Rev. Lett.}\ }\textbf {\bibinfo
  {volume} {119}},\ \bibinfo {pages} {176807} (\bibinfo {year}
  {2017})}\BibitemShut {NoStop}%
\bibitem [{\citenamefont {Jones}\ \emph {et~al.}(2018)\citenamefont {Jones},
  \citenamefont {Fogarty}, \citenamefont {Morello}, \citenamefont {Gyure},
  \citenamefont {Dzurak},\ and\ \citenamefont {Ladd}}]{Jones2018}%
  \BibitemOpen
  \bibfield  {author} {\bibinfo {author} {\bibfnamefont {C.}~\bibnamefont
  {Jones}}, \bibinfo {author} {\bibfnamefont {M.~A.}\ \bibnamefont {Fogarty}},
  \bibinfo {author} {\bibfnamefont {A.}~\bibnamefont {Morello}}, \bibinfo
  {author} {\bibfnamefont {M.~F.}\ \bibnamefont {Gyure}}, \bibinfo {author}
  {\bibfnamefont {A.~S.}\ \bibnamefont {Dzurak}}, \ and\ \bibinfo {author}
  {\bibfnamefont {T.~D.}\ \bibnamefont {Ladd}},\ }\href@noop {} {\bibfield
  {journal} {\bibinfo  {journal} {Phys. Rev. X}\ }\textbf {\bibinfo {volume}
  {8}},\ \bibinfo {pages} {021058} (\bibinfo {year} {2018})}\BibitemShut
  {NoStop}%
\bibitem [{\citenamefont {Fogarty}\ \emph {et~al.}(2018)\citenamefont
  {Fogarty}, \citenamefont {Chan}, \citenamefont {Hensen}, \citenamefont
  {Huang}, \citenamefont {Tanttu}, \citenamefont {Yang}, \citenamefont
  {Laucht}, \citenamefont {Veldhorst}, \citenamefont {Hudson}, \citenamefont
  {Itoh}, \citenamefont {Culcer}, \citenamefont {Ladd}, \citenamefont
  {Morello},\ and\ \citenamefont {Dzurak}}]{Fogarty2018}%
  \BibitemOpen
  \bibfield  {author} {\bibinfo {author} {\bibfnamefont {M.~A.}\ \bibnamefont
  {Fogarty}}, \bibinfo {author} {\bibfnamefont {K.~W.}\ \bibnamefont {Chan}},
  \bibinfo {author} {\bibfnamefont {B.}~\bibnamefont {Hensen}}, \bibinfo
  {author} {\bibfnamefont {W.}~\bibnamefont {Huang}}, \bibinfo {author}
  {\bibfnamefont {T.}~\bibnamefont {Tanttu}}, \bibinfo {author} {\bibfnamefont
  {C.~H.}\ \bibnamefont {Yang}}, \bibinfo {author} {\bibfnamefont
  {A.}~\bibnamefont {Laucht}}, \bibinfo {author} {\bibfnamefont
  {M.}~\bibnamefont {Veldhorst}}, \bibinfo {author} {\bibfnamefont {F.~E.}\
  \bibnamefont {Hudson}}, \bibinfo {author} {\bibfnamefont {K.~M.}\
  \bibnamefont {Itoh}}, \bibinfo {author} {\bibfnamefont {D.}~\bibnamefont
  {Culcer}}, \bibinfo {author} {\bibfnamefont {T.~D.}\ \bibnamefont {Ladd}},
  \bibinfo {author} {\bibfnamefont {A.}~\bibnamefont {Morello}}, \ and\
  \bibinfo {author} {\bibfnamefont {A.~S.}\ \bibnamefont {Dzurak}},\
  }\href@noop {} {\bibfield  {journal} {\bibinfo  {journal} {Nat. Commun.}\
  }\textbf {\bibinfo {volume} {9}},\ \bibinfo {pages} {4370} (\bibinfo {year}
  {2018})}\BibitemShut {NoStop}%
\bibitem [{\citenamefont {Veldhorst}\ \emph {et~al.}(2017)\citenamefont
  {Veldhorst}, \citenamefont {Eenink}, \citenamefont {Yang},\ and\
  \citenamefont {Dzurak}}]{Veldhorst2017}%
  \BibitemOpen
  \bibfield  {author} {\bibinfo {author} {\bibfnamefont {M.}~\bibnamefont
  {Veldhorst}}, \bibinfo {author} {\bibfnamefont {H.~G.~J.}\ \bibnamefont
  {Eenink}}, \bibinfo {author} {\bibfnamefont {C.~H.}\ \bibnamefont {Yang}}, \
  and\ \bibinfo {author} {\bibfnamefont {A.~S.}\ \bibnamefont {Dzurak}},\
  }\href@noop {} {\bibfield  {journal} {\bibinfo  {journal} {Nat. Commun.}\
  }\textbf {\bibinfo {volume} {8}},\ \bibinfo {pages} {1766} (\bibinfo {year}
  {2017})}\BibitemShut {NoStop}%
\bibitem [{\citenamefont {Brauns}\ \emph {et~al.}(2016)\citenamefont {Brauns},
  \citenamefont {Ridderbos}, \citenamefont {Li}, \citenamefont {Bakkers},\ and\
  \citenamefont {Zwanenburg}}]{Brauns2016}%
  \BibitemOpen
  \bibfield  {author} {\bibinfo {author} {\bibfnamefont {M.}~\bibnamefont
  {Brauns}}, \bibinfo {author} {\bibfnamefont {J.}~\bibnamefont {Ridderbos}},
  \bibinfo {author} {\bibfnamefont {A.}~\bibnamefont {Li}}, \bibinfo {author}
  {\bibfnamefont {E.~P. A.~M.}\ \bibnamefont {Bakkers}}, \ and\ \bibinfo
  {author} {\bibfnamefont {F.~A.}\ \bibnamefont {Zwanenburg}},\ }\href@noop {}
  {\bibfield  {journal} {\bibinfo  {journal} {Phys. Rev. B}\ }\textbf {\bibinfo
  {volume} {93}},\ \bibinfo {pages} {121408} (\bibinfo {year}
  {2016})}\BibitemShut {NoStop}%
\bibitem [{\citenamefont {Stepanenko}\ \emph {et~al.}(2003)\citenamefont
  {Stepanenko}, \citenamefont {Bonesteel}, \citenamefont {DiVincenzo},
  \citenamefont {Burkard},\ and\ \citenamefont {Loss}}]{Stepanenko2003}%
  \BibitemOpen
  \bibfield  {author} {\bibinfo {author} {\bibfnamefont {D.}~\bibnamefont
  {Stepanenko}}, \bibinfo {author} {\bibfnamefont {N.~E.}\ \bibnamefont
  {Bonesteel}}, \bibinfo {author} {\bibfnamefont {D.~P.}\ \bibnamefont
  {DiVincenzo}}, \bibinfo {author} {\bibfnamefont {G.}~\bibnamefont {Burkard}},
  \ and\ \bibinfo {author} {\bibfnamefont {D.}~\bibnamefont {Loss}},\
  }\href@noop {} {\bibfield  {journal} {\bibinfo  {journal} {Phys. Rev. B}\
  }\textbf {\bibinfo {volume} {68}},\ \bibinfo {pages} {115306} (\bibinfo
  {year} {2003})}\BibitemShut {NoStop}%
\bibitem [{\citenamefont {Rashba}(1960)}]{Rashba1960}%
  \BibitemOpen
  \bibfield  {author} {\bibinfo {author} {\bibfnamefont {E.}~\bibnamefont
  {Rashba}},\ }\href@noop {} {\bibfield  {journal} {\bibinfo  {journal} {Sov.
  Phys. Solid State}\ }\textbf {\bibinfo {volume} {2}},\ \bibinfo {pages}
  {1109–1122} (\bibinfo {year} {1960})}\BibitemShut {NoStop}%
\bibitem [{\citenamefont {Dresselhaus}(1955)}]{Dresselhaus1955}%
  \BibitemOpen
  \bibfield  {author} {\bibinfo {author} {\bibfnamefont {G.}~\bibnamefont
  {Dresselhaus}},\ }\href@noop {} {\bibfield  {journal} {\bibinfo  {journal}
  {Phys. Rev.}\ }\textbf {\bibinfo {volume} {100}},\ \bibinfo {pages} {580}
  (\bibinfo {year} {1955})}\BibitemShut {NoStop}%
\bibitem [{\citenamefont {Golub}\ and\ \citenamefont
  {Ivchenko}(2004)}]{Golub2004}%
  \BibitemOpen
  \bibfield  {author} {\bibinfo {author} {\bibfnamefont {L.~E.}\ \bibnamefont
  {Golub}}\ and\ \bibinfo {author} {\bibfnamefont {E.~L.}\ \bibnamefont
  {Ivchenko}},\ }\href@noop {} {\bibfield  {journal} {\bibinfo  {journal}
  {Phys. Rev. B}\ }\textbf {\bibinfo {volume} {69}},\ \bibinfo {pages} {115333}
  (\bibinfo {year} {2004})}\BibitemShut {NoStop}%
\bibitem [{\citenamefont {Nestoklon}\ \emph {et~al.}(2008)\citenamefont
  {Nestoklon}, \citenamefont {Ivchenko}, \citenamefont {Jancu},\ and\
  \citenamefont {Voisin}}]{Nestoklon2008}%
  \BibitemOpen
  \bibfield  {author} {\bibinfo {author} {\bibfnamefont {M.~O.}\ \bibnamefont
  {Nestoklon}}, \bibinfo {author} {\bibfnamefont {E.~L.}\ \bibnamefont
  {Ivchenko}}, \bibinfo {author} {\bibfnamefont {J.-M.}\ \bibnamefont {Jancu}},
  \ and\ \bibinfo {author} {\bibfnamefont {P.}~\bibnamefont {Voisin}},\
  }\href@noop {} {\bibfield  {journal} {\bibinfo  {journal} {Phys. Rev. B}\
  }\textbf {\bibinfo {volume} {77}},\ \bibinfo {pages} {155328} (\bibinfo
  {year} {2008})}\BibitemShut {NoStop}%
\bibitem [{\citenamefont {Crippa}\ \emph {et~al.}(2018)\citenamefont {Crippa},
  \citenamefont {Maurand}, \citenamefont {Bourdet}, \citenamefont
  {Kotekar-Patil}, \citenamefont {Amisse}, \citenamefont {Jehl}, \citenamefont
  {Sanquer}, \citenamefont {Lavi\'eville}, \citenamefont {Bohuslavskyi},
  \citenamefont {Hutin}, \citenamefont {Barraud}, \citenamefont {Vinet},
  \citenamefont {Niquet},\ and\ \citenamefont {De~Franceschi}}]{Crippa2018}%
  \BibitemOpen
  \bibfield  {author} {\bibinfo {author} {\bibfnamefont {A.}~\bibnamefont
  {Crippa}}, \bibinfo {author} {\bibfnamefont {R.}~\bibnamefont {Maurand}},
  \bibinfo {author} {\bibfnamefont {L.}~\bibnamefont {Bourdet}}, \bibinfo
  {author} {\bibfnamefont {D.}~\bibnamefont {Kotekar-Patil}}, \bibinfo {author}
  {\bibfnamefont {A.}~\bibnamefont {Amisse}}, \bibinfo {author} {\bibfnamefont
  {X.}~\bibnamefont {Jehl}}, \bibinfo {author} {\bibfnamefont {M.}~\bibnamefont
  {Sanquer}}, \bibinfo {author} {\bibfnamefont {R.}~\bibnamefont
  {Lavi\'eville}}, \bibinfo {author} {\bibfnamefont {H.}~\bibnamefont
  {Bohuslavskyi}}, \bibinfo {author} {\bibfnamefont {L.}~\bibnamefont {Hutin}},
  \bibinfo {author} {\bibfnamefont {S.}~\bibnamefont {Barraud}}, \bibinfo
  {author} {\bibfnamefont {M.}~\bibnamefont {Vinet}}, \bibinfo {author}
  {\bibfnamefont {Y.-M.}\ \bibnamefont {Niquet}}, \ and\ \bibinfo {author}
  {\bibfnamefont {S.}~\bibnamefont {De~Franceschi}},\ }\href@noop {} {\bibfield
   {journal} {\bibinfo  {journal} {Phys. Rev. Lett.}\ }\textbf {\bibinfo
  {volume} {120}},\ \bibinfo {pages} {137702} (\bibinfo {year}
  {2018})}\BibitemShut {NoStop}%
\bibitem [{\citenamefont {Yang}\ \emph {et~al.}(2013)\citenamefont {Yang},
  \citenamefont {Rossi}, \citenamefont {Ruskov}, \citenamefont {Lai},
  \citenamefont {Mohiyaddin}, \citenamefont {Lee}, \citenamefont {Tahan},
  \citenamefont {Klimeck}, \citenamefont {Morello},\ and\ \citenamefont
  {Dzurak}}]{Yang2013}%
  \BibitemOpen
  \bibfield  {author} {\bibinfo {author} {\bibfnamefont {C.~H.}\ \bibnamefont
  {Yang}}, \bibinfo {author} {\bibfnamefont {A.}~\bibnamefont {Rossi}},
  \bibinfo {author} {\bibfnamefont {R.}~\bibnamefont {Ruskov}}, \bibinfo
  {author} {\bibfnamefont {N.~S.}\ \bibnamefont {Lai}}, \bibinfo {author}
  {\bibfnamefont {F.~A.}\ \bibnamefont {Mohiyaddin}}, \bibinfo {author}
  {\bibfnamefont {S.}~\bibnamefont {Lee}}, \bibinfo {author} {\bibfnamefont
  {C.}~\bibnamefont {Tahan}}, \bibinfo {author} {\bibfnamefont
  {G.}~\bibnamefont {Klimeck}}, \bibinfo {author} {\bibfnamefont
  {A.}~\bibnamefont {Morello}}, \ and\ \bibinfo {author} {\bibfnamefont
  {A.~S.}\ \bibnamefont {Dzurak}},\ }\href@noop {} {\bibfield  {journal}
  {\bibinfo  {journal} {Nat. Commun.}\ }\textbf {\bibinfo {volume} {4}},\
  \bibinfo {pages} {2069} (\bibinfo {year} {2013})},\ \bibinfo {note}
  {article}\BibitemShut {NoStop}%
\bibitem [{\citenamefont {Pla}\ \emph {et~al.}(2018)\citenamefont {Pla},
  \citenamefont {Bienfait}, \citenamefont {Pica}, \citenamefont {Mansir},
  \citenamefont {Mohiyaddin}, \citenamefont {Zeng}, \citenamefont {Niquet},
  \citenamefont {Morello}, \citenamefont {Schenkel}, \citenamefont {Morton},\
  and\ \citenamefont {Bertet}}]{Pla2018}%
  \BibitemOpen
  \bibfield  {author} {\bibinfo {author} {\bibfnamefont {J.~J.}\ \bibnamefont
  {Pla}}, \bibinfo {author} {\bibfnamefont {A.}~\bibnamefont {Bienfait}},
  \bibinfo {author} {\bibfnamefont {G.}~\bibnamefont {Pica}}, \bibinfo {author}
  {\bibfnamefont {J.}~\bibnamefont {Mansir}}, \bibinfo {author} {\bibfnamefont
  {F.~A.}\ \bibnamefont {Mohiyaddin}}, \bibinfo {author} {\bibfnamefont
  {Z.}~\bibnamefont {Zeng}}, \bibinfo {author} {\bibfnamefont {Y.~M.}\
  \bibnamefont {Niquet}}, \bibinfo {author} {\bibfnamefont {A.}~\bibnamefont
  {Morello}}, \bibinfo {author} {\bibfnamefont {T.}~\bibnamefont {Schenkel}},
  \bibinfo {author} {\bibfnamefont {J.~J.~L.}\ \bibnamefont {Morton}}, \ and\
  \bibinfo {author} {\bibfnamefont {P.}~\bibnamefont {Bertet}},\ }\href@noop {}
  {\bibfield  {journal} {\bibinfo  {journal} {Phys. Rev. Appl.}\ }\textbf
  {\bibinfo {volume} {9}},\ \bibinfo {pages} {044014} (\bibinfo {year}
  {2018})}\BibitemShut {NoStop}%
\bibitem [{\citenamefont {Roth}(1960)}]{Roth1960}%
  \BibitemOpen
  \bibfield  {author} {\bibinfo {author} {\bibfnamefont {L.~M.}\ \bibnamefont
  {Roth}},\ }\href@noop {} {\bibfield  {journal} {\bibinfo  {journal} {Phys.
  Rev.}\ }\textbf {\bibinfo {volume} {118}},\ \bibinfo {pages} {1534} (\bibinfo
  {year} {1960})}\BibitemShut {NoStop}%
\bibitem [{\citenamefont {Muhonen}\ \emph {et~al.}(2014)\citenamefont
  {Muhonen}, \citenamefont {Dehollain}, \citenamefont {Laucht}, \citenamefont
  {Hudson}, \citenamefont {Kalra}, \citenamefont {Sekiguchi}, \citenamefont
  {Itoh}, \citenamefont {Jamieson}, \citenamefont {McCallum}, \citenamefont
  {Dzurak},\ and\ \citenamefont {Morello}}]{Muhonen2014}%
  \BibitemOpen
  \bibfield  {author} {\bibinfo {author} {\bibfnamefont {J.~T.}\ \bibnamefont
  {Muhonen}}, \bibinfo {author} {\bibfnamefont {J.~P.}\ \bibnamefont
  {Dehollain}}, \bibinfo {author} {\bibfnamefont {A.}~\bibnamefont {Laucht}},
  \bibinfo {author} {\bibfnamefont {F.~E.}\ \bibnamefont {Hudson}}, \bibinfo
  {author} {\bibfnamefont {R.}~\bibnamefont {Kalra}}, \bibinfo {author}
  {\bibfnamefont {T.}~\bibnamefont {Sekiguchi}}, \bibinfo {author}
  {\bibfnamefont {K.~M.}\ \bibnamefont {Itoh}}, \bibinfo {author}
  {\bibfnamefont {D.~N.}\ \bibnamefont {Jamieson}}, \bibinfo {author}
  {\bibfnamefont {J.~C.}\ \bibnamefont {McCallum}}, \bibinfo {author}
  {\bibfnamefont {A.~S.}\ \bibnamefont {Dzurak}}, \ and\ \bibinfo {author}
  {\bibfnamefont {A.}~\bibnamefont {Morello}},\ }\href@noop {} {\bibfield
  {journal} {\bibinfo  {journal} {Nat. Nanotechnol.}\ }\textbf {\bibinfo
  {volume} {9}},\ \bibinfo {pages} {986} (\bibinfo {year} {2014})}\BibitemShut
  {NoStop}%
\bibitem [{\citenamefont {Elzerman}\ \emph {et~al.}(2004)\citenamefont
  {Elzerman}, \citenamefont {Hanson}, \citenamefont {Willems~van Beveren},
  \citenamefont {Witkamp}, \citenamefont {Vandersypen},\ and\ \citenamefont
  {Kouwenhoven}}]{Elzerman2004}%
  \BibitemOpen
  \bibfield  {author} {\bibinfo {author} {\bibfnamefont {J.~M.}\ \bibnamefont
  {Elzerman}}, \bibinfo {author} {\bibfnamefont {R.}~\bibnamefont {Hanson}},
  \bibinfo {author} {\bibfnamefont {L.~H.}\ \bibnamefont {Willems~van
  Beveren}}, \bibinfo {author} {\bibfnamefont {B.}~\bibnamefont {Witkamp}},
  \bibinfo {author} {\bibfnamefont {L.~M.~K.}\ \bibnamefont {Vandersypen}}, \
  and\ \bibinfo {author} {\bibfnamefont {L.~P.}\ \bibnamefont {Kouwenhoven}},\
  }\href@noop {} {\bibfield  {journal} {\bibinfo  {journal} {Nature}\ }\textbf
  {\bibinfo {volume} {430}},\ \bibinfo {pages} {431} (\bibinfo {year}
  {2004})}\BibitemShut {NoStop}%
\bibitem [{\citenamefont {Feher}(1956)}]{Feher1956}%
  \BibitemOpen
  \bibfield  {author} {\bibinfo {author} {\bibfnamefont {G.}~\bibnamefont
  {Feher}},\ }\href@noop {} {\bibfield  {journal} {\bibinfo  {journal} {Phys.
  Rev.}\ }\textbf {\bibinfo {volume} {103}},\ \bibinfo {pages} {834} (\bibinfo
  {year} {1956})}\BibitemShut {NoStop}%
\bibitem [{\citenamefont {Veldhorst}\ \emph {et~al.}(2014)\citenamefont
  {Veldhorst}, \citenamefont {Hwang}, \citenamefont {Yang}, \citenamefont
  {Leenstra}, \citenamefont {de~Ronde}, \citenamefont {Dehollain},
  \citenamefont {Muhonen}, \citenamefont {Hudson}, \citenamefont {Itoh},
  \citenamefont {Morello},\ and\ \citenamefont {Dzurak}}]{Veldhorst2014}%
  \BibitemOpen
  \bibfield  {author} {\bibinfo {author} {\bibfnamefont {M.}~\bibnamefont
  {Veldhorst}}, \bibinfo {author} {\bibfnamefont {J.~C.~C.}\ \bibnamefont
  {Hwang}}, \bibinfo {author} {\bibfnamefont {C.~H.}\ \bibnamefont {Yang}},
  \bibinfo {author} {\bibfnamefont {A.~W.}\ \bibnamefont {Leenstra}}, \bibinfo
  {author} {\bibfnamefont {B.}~\bibnamefont {de~Ronde}}, \bibinfo {author}
  {\bibfnamefont {J.~P.}\ \bibnamefont {Dehollain}}, \bibinfo {author}
  {\bibfnamefont {J.~T.}\ \bibnamefont {Muhonen}}, \bibinfo {author}
  {\bibfnamefont {F.~E.}\ \bibnamefont {Hudson}}, \bibinfo {author}
  {\bibfnamefont {K.~M.}\ \bibnamefont {Itoh}}, \bibinfo {author}
  {\bibfnamefont {A.}~\bibnamefont {Morello}}, \ and\ \bibinfo {author}
  {\bibfnamefont {A.~S.}\ \bibnamefont {Dzurak}},\ }\href@noop {} {\bibfield
  {journal} {\bibinfo  {journal} {Nat. Nanotechnol.}\ }\textbf {\bibinfo
  {volume} {9}},\ \bibinfo {pages} {981} (\bibinfo {year} {2014})}\BibitemShut
  {NoStop}%
\bibitem [{\citenamefont {Chan}\ \emph {et~al.}(2018)\citenamefont {Chan},
  \citenamefont {Huang}, \citenamefont {Yang}, \citenamefont {Hwang},
  \citenamefont {Hensen}, \citenamefont {Tanttu}, \citenamefont {Hudson},
  \citenamefont {Itoh}, \citenamefont {Laucht}, \citenamefont {Morello},\ and\
  \citenamefont {Dzurak}}]{Chan2018}%
  \BibitemOpen
  \bibfield  {author} {\bibinfo {author} {\bibfnamefont {K.~W.}\ \bibnamefont
  {Chan}}, \bibinfo {author} {\bibfnamefont {W.}~\bibnamefont {Huang}},
  \bibinfo {author} {\bibfnamefont {C.~H.}\ \bibnamefont {Yang}}, \bibinfo
  {author} {\bibfnamefont {J.~C.~C.}\ \bibnamefont {Hwang}}, \bibinfo {author}
  {\bibfnamefont {B.}~\bibnamefont {Hensen}}, \bibinfo {author} {\bibfnamefont
  {T.}~\bibnamefont {Tanttu}}, \bibinfo {author} {\bibfnamefont {F.~E.}\
  \bibnamefont {Hudson}}, \bibinfo {author} {\bibfnamefont {K.~M.}\
  \bibnamefont {Itoh}}, \bibinfo {author} {\bibfnamefont {A.}~\bibnamefont
  {Laucht}}, \bibinfo {author} {\bibfnamefont {A.}~\bibnamefont {Morello}}, \
  and\ \bibinfo {author} {\bibfnamefont {A.~S.}\ \bibnamefont {Dzurak}},\
  }\href@noop {} {\bibfield  {journal} {\bibinfo  {journal} {Phys. Rev.
  Applied}\ }\textbf {\bibinfo {volume} {10}},\ \bibinfo {pages} {044017}
  (\bibinfo {year} {2018})}\BibitemShut {NoStop}%
\bibitem [{\citenamefont {Taylor}\ \emph {et~al.}(2005)\citenamefont {Taylor},
  \citenamefont {Engel}, \citenamefont {D{\"u}r}, \citenamefont {Yacoby},
  \citenamefont {Marcus}, \citenamefont {Zoller},\ and\ \citenamefont
  {Lukin}}]{Taylor2005}%
  \BibitemOpen
  \bibfield  {author} {\bibinfo {author} {\bibfnamefont {J.~M.}\ \bibnamefont
  {Taylor}}, \bibinfo {author} {\bibfnamefont {H.-A.}\ \bibnamefont {Engel}},
  \bibinfo {author} {\bibfnamefont {W.}~\bibnamefont {D{\"u}r}}, \bibinfo
  {author} {\bibfnamefont {A.}~\bibnamefont {Yacoby}}, \bibinfo {author}
  {\bibfnamefont {C.~M.}\ \bibnamefont {Marcus}}, \bibinfo {author}
  {\bibfnamefont {P.}~\bibnamefont {Zoller}}, \ and\ \bibinfo {author}
  {\bibfnamefont {M.~D.}\ \bibnamefont {Lukin}},\ }\href@noop {} {\bibfield
  {journal} {\bibinfo  {journal} {Nat. Phys.}\ }\textbf {\bibinfo {volume}
  {1}},\ \bibinfo {pages} {177} (\bibinfo {year} {2005})},\ \bibinfo {note}
  {article}\BibitemShut {NoStop}%
\bibitem [{\citenamefont {Scarlino}\ \emph {et~al.}(2014)\citenamefont
  {Scarlino}, \citenamefont {Kawakami}, \citenamefont {Stano}, \citenamefont
  {Shafiei}, \citenamefont {Reichl}, \citenamefont {Wegscheider},\ and\
  \citenamefont {Vandersypen}}]{Scarlino2014}%
  \BibitemOpen
  \bibfield  {author} {\bibinfo {author} {\bibfnamefont {P.}~\bibnamefont
  {Scarlino}}, \bibinfo {author} {\bibfnamefont {E.}~\bibnamefont {Kawakami}},
  \bibinfo {author} {\bibfnamefont {P.}~\bibnamefont {Stano}}, \bibinfo
  {author} {\bibfnamefont {M.}~\bibnamefont {Shafiei}}, \bibinfo {author}
  {\bibfnamefont {C.}~\bibnamefont {Reichl}}, \bibinfo {author} {\bibfnamefont
  {W.}~\bibnamefont {Wegscheider}}, \ and\ \bibinfo {author} {\bibfnamefont
  {L.~M.~K.}\ \bibnamefont {Vandersypen}},\ }\href@noop {} {\bibfield
  {journal} {\bibinfo  {journal} {Phys. Rev. Lett.}\ }\textbf {\bibinfo
  {volume} {113}},\ \bibinfo {pages} {256802} (\bibinfo {year}
  {2014})}\BibitemShut {NoStop}%
\bibitem [{\citenamefont {Taylor}\ \emph {et~al.}(2007)\citenamefont {Taylor},
  \citenamefont {Petta}, \citenamefont {Johnson}, \citenamefont {Yacoby},
  \citenamefont {Marcus},\ and\ \citenamefont {Lukin}}]{Taylor2007}%
  \BibitemOpen
  \bibfield  {author} {\bibinfo {author} {\bibfnamefont {J.~M.}\ \bibnamefont
  {Taylor}}, \bibinfo {author} {\bibfnamefont {J.~R.}\ \bibnamefont {Petta}},
  \bibinfo {author} {\bibfnamefont {A.~C.}\ \bibnamefont {Johnson}}, \bibinfo
  {author} {\bibfnamefont {A.}~\bibnamefont {Yacoby}}, \bibinfo {author}
  {\bibfnamefont {C.~M.}\ \bibnamefont {Marcus}}, \ and\ \bibinfo {author}
  {\bibfnamefont {M.~D.}\ \bibnamefont {Lukin}},\ }\href@noop {} {\bibfield
  {journal} {\bibinfo  {journal} {Phys. Rev. B}\ }\textbf {\bibinfo {volume}
  {76}},\ \bibinfo {pages} {035315} (\bibinfo {year} {2007})}\BibitemShut
  {NoStop}%
\bibitem [{\citenamefont {Underwood}(2017)}]{Underwood2017}%
  \BibitemOpen
  \bibfield  {author} {\bibinfo {author} {\bibfnamefont {D.}~\bibnamefont
  {Underwood}},\ }\href@noop {} {\bibfield  {journal} {\bibinfo  {journal}
  {bulletin of the american physical society}\ } (\bibinfo {year}
  {2017})}\BibitemShut {NoStop}%
\bibitem [{\citenamefont {Eng}\ \emph {et~al.}(2015)\citenamefont {Eng},
  \citenamefont {Ladd}, \citenamefont {Smith}, \citenamefont {Borselli},
  \citenamefont {Kiselev}, \citenamefont {Fong}, \citenamefont {Holabird},
  \citenamefont {Hazard}, \citenamefont {Huang}, \citenamefont {Deelman},
  \citenamefont {Milosavljevic}, \citenamefont {Schmitz}, \citenamefont {Ross},
  \citenamefont {Gyure},\ and\ \citenamefont {Hunter}}]{Eng2015}%
  \BibitemOpen
  \bibfield  {author} {\bibinfo {author} {\bibfnamefont {K.}~\bibnamefont
  {Eng}}, \bibinfo {author} {\bibfnamefont {T.~D.}\ \bibnamefont {Ladd}},
  \bibinfo {author} {\bibfnamefont {A.}~\bibnamefont {Smith}}, \bibinfo
  {author} {\bibfnamefont {M.~G.}\ \bibnamefont {Borselli}}, \bibinfo {author}
  {\bibfnamefont {A.~A.}\ \bibnamefont {Kiselev}}, \bibinfo {author}
  {\bibfnamefont {B.~H.}\ \bibnamefont {Fong}}, \bibinfo {author}
  {\bibfnamefont {K.~S.}\ \bibnamefont {Holabird}}, \bibinfo {author}
  {\bibfnamefont {T.~M.}\ \bibnamefont {Hazard}}, \bibinfo {author}
  {\bibfnamefont {B.}~\bibnamefont {Huang}}, \bibinfo {author} {\bibfnamefont
  {P.~W.}\ \bibnamefont {Deelman}}, \bibinfo {author} {\bibfnamefont
  {I.}~\bibnamefont {Milosavljevic}}, \bibinfo {author} {\bibfnamefont {A.~E.}\
  \bibnamefont {Schmitz}}, \bibinfo {author} {\bibfnamefont {R.~S.}\
  \bibnamefont {Ross}}, \bibinfo {author} {\bibfnamefont {M.~F.}\ \bibnamefont
  {Gyure}}, \ and\ \bibinfo {author} {\bibfnamefont {A.~T.}\ \bibnamefont
  {Hunter}},\ }\href {\doibase 10.1126/sciadv.1500214} {\bibfield  {journal}
  {\bibinfo  {journal} {Sci. Adv.}\ }\textbf {\bibinfo {volume} {1}} (\bibinfo
  {year} {2015}),\ 10.1126/sciadv.1500214}\BibitemShut {NoStop}%
\bibitem [{\citenamefont {Harvey-Collard}\ \emph {et~al.}(2018)\citenamefont
  {Harvey-Collard}, \citenamefont {D'Anjou}, \citenamefont {Rudolph},
  \citenamefont {Jacobson}, \citenamefont {Dominguez}, \citenamefont
  {Ten~Eyck}, \citenamefont {Wendt}, \citenamefont {Pluym}, \citenamefont
  {Lilly}, \citenamefont {Coish}, \citenamefont {Pioro-Ladri\`ere},\ and\
  \citenamefont {Carroll}}]{Harvey-Collard2018}%
  \BibitemOpen
  \bibfield  {author} {\bibinfo {author} {\bibfnamefont {P.}~\bibnamefont
  {Harvey-Collard}}, \bibinfo {author} {\bibfnamefont {B.}~\bibnamefont
  {D'Anjou}}, \bibinfo {author} {\bibfnamefont {M.}~\bibnamefont {Rudolph}},
  \bibinfo {author} {\bibfnamefont {N.~T.}\ \bibnamefont {Jacobson}}, \bibinfo
  {author} {\bibfnamefont {J.}~\bibnamefont {Dominguez}}, \bibinfo {author}
  {\bibfnamefont {G.~A.}\ \bibnamefont {Ten~Eyck}}, \bibinfo {author}
  {\bibfnamefont {J.~R.}\ \bibnamefont {Wendt}}, \bibinfo {author}
  {\bibfnamefont {T.}~\bibnamefont {Pluym}}, \bibinfo {author} {\bibfnamefont
  {M.~P.}\ \bibnamefont {Lilly}}, \bibinfo {author} {\bibfnamefont {W.~A.}\
  \bibnamefont {Coish}}, \bibinfo {author} {\bibfnamefont {M.}~\bibnamefont
  {Pioro-Ladri\`ere}}, \ and\ \bibinfo {author} {\bibfnamefont {M.~S.}\
  \bibnamefont {Carroll}},\ }\href@noop {} {\bibfield  {journal} {\bibinfo
  {journal} {Phys. Rev. X}\ }\textbf {\bibinfo {volume} {8}},\ \bibinfo {pages}
  {021046} (\bibinfo {year} {2018})}\BibitemShut {NoStop}%
\bibitem [{\citenamefont {Shevchenko}\ \emph {et~al.}(2010)\citenamefont
  {Shevchenko}, \citenamefont {Ashhab},\ and\ \citenamefont
  {Nori}}]{Shevchenko2010}%
  \BibitemOpen
  \bibfield  {author} {\bibinfo {author} {\bibfnamefont {S.~N.}\ \bibnamefont
  {Shevchenko}}, \bibinfo {author} {\bibfnamefont {S.}~\bibnamefont {Ashhab}},
  \ and\ \bibinfo {author} {\bibfnamefont {F.}~\bibnamefont {Nori}},\
  }\href@noop {} {\bibfield  {journal} {\bibinfo  {journal} {Phys. Rep.}\
  }\textbf {\bibinfo {volume} {492}},\ \bibinfo {pages} {1 } (\bibinfo {year}
  {2010})}\BibitemShut {NoStop}%
\bibitem [{\citenamefont {Nichol}\ \emph {et~al.}(2015)\citenamefont {Nichol},
  \citenamefont {Harvey}, \citenamefont {Shulman}, \citenamefont {Pal},
  \citenamefont {Umansky}, \citenamefont {Rashba}, \citenamefont {Halperin},\
  and\ \citenamefont {Yacoby}}]{Nichol2015}%
  \BibitemOpen
  \bibfield  {author} {\bibinfo {author} {\bibfnamefont {J.~M.}\ \bibnamefont
  {Nichol}}, \bibinfo {author} {\bibfnamefont {S.~P.}\ \bibnamefont {Harvey}},
  \bibinfo {author} {\bibfnamefont {M.~D.}\ \bibnamefont {Shulman}}, \bibinfo
  {author} {\bibfnamefont {A.}~\bibnamefont {Pal}}, \bibinfo {author}
  {\bibfnamefont {V.}~\bibnamefont {Umansky}}, \bibinfo {author} {\bibfnamefont
  {E.~I.}\ \bibnamefont {Rashba}}, \bibinfo {author} {\bibfnamefont {B.~I.}\
  \bibnamefont {Halperin}}, \ and\ \bibinfo {author} {\bibfnamefont
  {A.}~\bibnamefont {Yacoby}},\ }\href@noop {} {\bibfield  {journal} {\bibinfo
  {journal} {Nat. Commun.}\ }\textbf {\bibinfo {volume} {6}},\ \bibinfo {pages}
  {7682} (\bibinfo {year} {2015})},\ \bibinfo {note} {article}\BibitemShut
  {NoStop}%
\end{thebibliography}%

\end{document}